\documentclass[12pt,reqno]{amsart}
\usepackage{main}
\usepackage{setspace}
\usepackage{booktabs,array,makecell,tabularx,multirow,adjustbox,pdflscape,longtable}
\usepackage{placeins}
\usepackage{titletoc}
\usepackage[style=apa,backend=biber,natbib=true,backref=true]{biblatex}
\usepackage{xurl}
\DefineBibliographyStrings{english}{backrefpage={},backrefpages={}}
\renewbibmacro*{pageref}{%
  \iflistundef{pageref}{}%
  {\printtext[brackets]{\printlist[pageref][-\value{listtotal}]{pageref}}}}
\newenvironment{tablebody}{%
  \begingroup
  \normalsize\setstretch{1}%
  \setlength{\tabcolsep}{6pt}%
}{\par\endgroup}

\graphicspath{{figures/}}
\renewcommand{\theHequation}{\thesection.\arabic{equation}}
\renewcommand{\theHfigure}{\thesection.\arabic{figure}}
\renewcommand{\theHtable}{\thesection.\arabic{table}}

\title{\Large The Geometry of Informativeness: A Bregman-Centroid Diagnostic for Partial Pooling in Multisite Trials}
\author{JoonHo Lee}
\date{\footnotesize October 2026. \\[0.5em]
Lee: College of Education, The University of Alabama, Tuscaloosa, AL, USA.
\href{mailto:jlee296@ua.edu}{jlee296@ua.edu}.
\\[0.5em]
This work was supported by the Institute of Education Sciences, U.S. Department of Education, through Grant R305D240078 to the University of Alabama. The opinions expressed are those of the author and do not represent views of the Institute or the U.S. Department of Education.
}

\begin{document}
\begin{abstract}
Two multisite trials can have the same informativeness yet differ in how reliably individual site effects can be estimated. We show that the informativeness index averages site-specific reliabilities on the logit scale and is a Bregman centroid under Bernoulli--Kullback--Leibler divergence. This explains the geometric mean of sampling variances and yields a measure of profile dispersion. Under stated conditions, we relate profile dispersion to estimation error and show when balanced allocation maximizes informativeness at fixed total enrollment. Simulations show that estimator performance depends on the reliability profile and whether the target is individual effects, ranks, or their distribution. An online credit recovery trial illustrates the diagnostics and their conditional posterior uncertainty. We recommend reporting informativeness with arithmetic mean reliability and profile dispersion. The accompanying R package \textsf{poolinfo} provides tools for computing the diagnostics, posterior summaries, and design calculations.

\end{abstract}
\maketitle
\pagestyle{plain}
\newpage

\section{Introduction}\label{sec:introduction}

Multisite randomized trials allow education researchers to ask not only whether an intervention works on average, but also how its effects vary across schools or districts \citep{Raudenbush2015}. These questions require different information. An estimate of the overall effect does not tell us how well the trial can distinguish effects at individual sites. Researchers who wish to identify settings with weak effects, rank sites, or describe the distribution of effects must separate genuine differences from sampling error \citep{Louis1984,ShenLouis1998,LeeEtAl2025}. A summary of how much information the trial provides for these tasks is therefore useful, particularly when site sizes and sampling variances differ.

Psychometric reliability provides a familiar starting point. A reliability coefficient compares true-score variation with total observed variation; its interpretation depends on what is being measured and what counts as error \citep{Cronbach1972,Brennan2001}. For a group mean, the relevant error variance is the sampling variance of that mean, so groups of different sizes can have different reliabilities \citep{SnijdersBosker2012}. For a site-effect estimate, the analogous comparison is between genuine variation in effects and sampling error. At fixed sampling variances, greater between-site variation increases reliability. This interpretation is consistent with the classical expression of reliability odds as a signal-to-noise ratio \citep{CronbachGleser1964} and with a conditional-projection definition of reliability \citep{Zumbo2026a}.

Meta-analysis uses a related variance comparison for a different purpose. The $I^2$ measure of \citet{HigginsThompson2002} quantifies variation attributable to heterogeneity relative to sampling error. With unequal within-study variances, their construction uses a representative sampling variance derived from inverse-variance weights. This representative variance need not equal either the arithmetic or the harmonic mean of the sampling variances \citep{Bohning2017}. Thus the same contrast between genuine variation and sampling error can describe heterogeneity or reliability, but an overall measure also depends on how unequal sampling variances are combined.

\citet{Crippa2016} take a different approach. They motivate $R_b$ by comparing the variance of a pooled random-effects estimate with the variance it would have if the study effects were measured without sampling error. The resulting expression is the arithmetic mean of the study-specific proportions of variance attributable to heterogeneity. For the same variance inputs, this equally weighted average of the individual proportions is also arithmetic mean reliability. This arithmetic mean of variance ratios generally differs from a ratio based on the arithmetic mean sampling variance. Reliability generalization makes another aspect of this choice explicit: when combining estimated coefficient alpha across studies, researchers choose transformations and weights to reflect the sampling properties of those coefficients \citep{RodriguezMaeda2006,LopezIbanez2024}.

For multisite trials, \citet{LeeEtAl2025} use a different summary to compare prior models and posterior estimators of site effects. Their informativeness index, $I$, relates between-site variance to the \emph{geometric} mean of the sampling variances. The geometric mean specifies a different averaging rule from the representative sampling variance in $I^2$ or an arithmetic average of individual reliabilities. It also leaves an interpretive question: in what sense is an index constructed from geometric mean sampling variance an average reliability? For the same variance inputs, these summaries agree when all sites have the same sampling variance; they can differ when precision varies. Clarifying this choice helps determine what comparisons of informativeness across trials or designs mean.

A second question concerns what any such average leaves out. Two trials can have the same informativeness even though one has similar precision across sites and the other contains both precise and imprecise estimates. The overall level of information does not describe this variation or show whether most reliability is concentrated in a few sites. These differences matter because estimating individual effects, ranking sites, and estimating their distribution are distinct tasks. Empirical Bayes methods already distinguish posterior summaries for these goals \citep{Louis1984,Ghosh1992,ShenLouis1998}. An information summary used to interpret their performance should therefore be examined together with the profile of site reliabilities.

In this paper, we show that informativeness averages site reliabilities on the logit scale. The logit is the log signal-to-noise ratio, which explains the geometric mean of sampling variances. We consider a finite collection of sites with equal weights, a common positive signal variance, and fixed positive sampling variances. Under conditional moment assumptions, the signal-to-total-variance ratio is the estimate's projection reliability. In the normal--normal model with known hyperparameters, it is also the weight on the site's estimate in the posterior mean. Using established Bregman-centroid results \citep{Bregman1967,Banerjee2005,NielsenNock2009}, we identify $I$ as the first-argument Bernoulli--Kullback--Leibler centroid of these reliabilities. Bregman projections have also been used to describe nonlinear measurement in psychometrics \citep{Zumbo2026b}. We choose the divergence to recover the existing index; boundedness of reliabilities alone does not determine this choice.

We then relate the reliability profile to estimation error. Arithmetic mean reliability determines the minimum integrated MSEL among affine estimators based on a site's own estimate, when the common mean, effect variance, and sampling variances are known. The dispersion $H$ measures the minimum discrepancy when the profile is replaced by a constant under the loss defining $I$. An effective site count, $J_{\mathrm{eff}}$, describes how evenly reliability is distributed across sites. The difference between arithmetic mean reliability and informativeness equals the rate of change of $H$ with the mean logit, holding centered logits fixed. A local approximation has a remainder that depends on profile shape. The difference can have either sign; the geometry does not make either average universally preferable.

We use these results to study allocation under fixed resources and to compare specified estimation procedures in simulations. We also report posterior summaries of the diagnostics, separating uncertainty conditional on the supplied first-stage variances from sensitivity to their estimation. An application to the Online Credit Recovery Study illustrates how to report the information level, dispersion, and concentration together. These summaries help interpret what a trial can reveal about its sites, but provide neither universal diagnostic cutoffs nor confidence guarantees for individual ranks.

Section~\ref{sec:setting} defines the inferential goals and reviews related indices. Sections~\ref{sec:reliability}--\ref{sec:dispersion} develop the reliability interpretation, geometry, and diagnostics; Section~\ref{sec:design} considers allocation. We present the simulations and application in Sections~\ref{sec:simulation} and~\ref{sec:ocrs}, followed by discussion. Proofs and methodological details are in Appendices A--G.

\section{Site-Specific Inference and the Informativeness Index}\label{sec:setting}

\subsection{The Multisite Model}\label{subsec:model}

We consider a two-stage model for $J$ site effects \citep{Rubin1981}. The first stage provides estimates $\hat{\tau}_j$ of $\tau_j$ and squared standard errors $v_j = \widehat{se}_j^2$. We treat these variances as known. Site estimates are independent conditional on the effects, and the effects are independent given $(\tau,\sigma^2)$:
\begin{equation}\label{eq:model}
\hat{\tau}_j \mid \tau_j \sim \Norm(\tau_j, v_j), \qquad
\tau_j \mid \tau, \sigma^2 \sim \Norm(\tau, \sigma^2), \qquad j = 1, \ldots, J,
\end{equation}
where $\tau$ and $\sigma^2$ are the mean and variance of site effects. A common planning approximation uses site sample size $n_j$, treated proportion $p_j$, equal within-arm outcome variances $s_j^2$, and residual variance reduction $R^2$:
\begin{equation}\label{eq:neyman}
v_j = \left(\frac{1}{p_j} + \frac{1}{1 - p_j}\right) \frac{s_j^2 (1 - R^2)}{n_j},
\end{equation}
so larger sites, balanced treatment allocation, and prognostic covariates can improve precision \citep{LeeEtAl2025}. Blocked or adjusted analyses require corresponding sampling variances and an explicit first-stage estimand. The average effect over persons in a site is $\tau_j^{\mathrm{person}}=N_j^{-1}\sum_{i\in j}\{Y_i(1)-Y_i(0)\}$, whereas a blocked regression can assign different weights to block effects. We use $\tau_j$ for the specified first-stage parameter and define it for the illustration in Section~\ref{sec:ocrs}.

Our targets are the finite vector $(\tau_1,\ldots,\tau_J)$, the ranks of its elements, and its uniformly weighted empirical distribution $G_J$ \citep{ShenLouis1998,LeeEtAl2025}. The hypermean $\tau$ is a parameter of the working distribution and is distinct from both the realized finite mean $J^{-1}\sum_j\tau_j$ and a trial effect weighted by the number of persons. Weights within sites are determined by the first-stage estimand, independently of the equal weighting of sites.

Conditional on the hyperparameters, the posterior for each site effect is normal,
\begin{equation}\label{eq:posterior}
\tau_j \mid \tau, \sigma^2, \hat{\tau}_j \sim \Norm\!\big(\tau_j^*, \; \sigma^2 (1 - S_j)\big), \qquad
\tau_j^* = \tau + (\hat{\tau}_j - \tau)\, S_j,
\end{equation}
with weight on the site estimate
\begin{equation}\label{eq:shrinkage}
S_j = \frac{\sigma^2}{\sigma^2 + v_j} \in (0, 1).
\end{equation}
When $v_j\ll\sigma^2$, the posterior mean is close to the site estimate, whereas $v_j\gg\sigma^2$ gives a posterior mean close to the hypermean. The conditional pooling weight $\omega_j = 1 - S_j$ is the coefficient on the hypermean in~\eqref{eq:posterior}. \citet{GelmanPardoe2006} instead compare posterior residual variances for a finite batch of effects at a specified model level. Their pooling measure need not equal $1-\bar S$, where $\bar S=J^{-1}\sum_jS_j$.

\subsection{Inferential Goals, Losses, and Posterior Summaries}\label{subsec:goals}

Following \citet{ShenLouis1998} and \citet{LeeEtAl2025}, we evaluate an estimate vector $d=(d_1,\ldots,d_J)$ under three losses that give equal weight to sites:
\begin{align*}
\mathrm{MSEL}&=J^{-1}\sum_j(d_j-\tau_j)^2,\\
\mathrm{MSELP}&=J^{-1}\sum_j(A_j/J-R_j/J)^2,\\
\mathrm{ISEL}&=\int\{F_d(t)-G_J(t)\}^2\,dt.
\end{align*}
Here $R_j$ and $A_j$ are the ascending ranks of effects and estimates, respectively, with average ranks for ties. The empirical distribution function (EDF) $F_d$ assigns equal mass to each estimate. RMSE is $\sqrt{\mathrm{MSEL}}$. Ranks are normalized as $R/J$, and the EDF loss concerns the realized finite collection of effects. A mixing distribution and $G_J$ are related but distinct targets. Under additional self-consistency conditions, a fitted mixing distribution may coincide with the average of its plug-in conditional CDFs \citep{Laird1978}.

Posterior means (PM) minimize posterior expected MSEL. We obtain PM by integrating over the hyperparameters held fixed in~\eqref{eq:posterior}. Constrained Bayes (CB) minimizes squared loss subject to constraints on the mean and spread of the finite ensemble \citep{Louis1984,Ghosh1992}. Exact matching of the centered spread requires posterior covariances. We use the approximation based on marginal variances in Appendix~\ref{subsec:F-fitting}.

The triple-goal summary (GR) assigns midpoint quantiles of the posterior expected finite EDF to sites ordered by posterior expected rank \citep{ShenLouis1998}. These quantiles minimize posterior expected ISEL among EDFs with $J$ equally weighted coordinates. Rank assignment labels the quantiles without changing the EDF. This result does not imply simultaneous unrestricted optimality under all three losses.

With unequal sampling variances, PM, unpooled estimates, and posterior expected ranks can order sites differently \citep{LairdLouis1989,Lockwood2002}. Expected ranks minimize squared-rank loss when arbitrary real scores are allowed. Restricting the actions to permutations or to ranks of effect estimates changes the optimization problem. Tie conventions are given in Appendix~\ref{subsec:F-fitting}. Confidence sets for ranks and selection require separate uncertainty procedures \citep{Mogstad2024}; profile diagnostics do not establish the certainty of individual ranks or selected lists.

\subsection{The Informativeness Index and Related Quantities}\label{subsec:informativeness}

To summarize the information that a trial provides about its site effects, \citet{LeeEtAl2025} define the informativeness index
\begin{equation}\label{eq:informativeness}
I = \frac{\sigma^2}{\sigma^2 + \tilde{v}}, \qquad
\tilde{v} = \exp\!\left(\frac{1}{J} \sum_{j=1}^{J} \ln v_j\right),
\end{equation}
which is the shrinkage weight for a hypothetical site whose sampling variance equals the geometric mean $\tilde{v}$ of the sampling variances. Their simulations related $I$ to performance for the three inferential goals. To assess their interpretation of $I$ as an average reliability, we establish why each $S_j$ is a reliability and how $I$ averages these coefficients.

Table~\ref{tab:crosswalk} compares $I$ with related summaries. The arithmetic mean $\bar S$ has a direct antecedent in the $R_b$ functional of \citet{Crippa2016}: $R_b=\bar S$ when both quantities use the same $\sigma^2,v_j$ and equal site weights. Different variance estimators or posterior averaging can give different estimates. Their motivation concerns variation in a pooled estimate; our losses concern site effects, ranks, and the finite EDF.

With unequal precisions, the representative sampling variance of \citet{HigginsThompson2002} for finite $J$ is
\begin{equation}\label{eq:typical-variance}
v_{\mathrm{typ}}=\frac{(J-1)W}{W^2-W_2},\qquad
W=\sum_jv_j^{-1},\quad W_2=\sum_jv_j^{-2},\quad J\ge2.
\end{equation}
The proportion obtained from fixed inputs, $R_{\mathrm{typ}}=\sigma^2/(\sigma^2+v_{\mathrm{typ}})$, differs from the truncated $Q$-based estimate $\widehat I_{\mathrm{HT}}^2=\max\{0,1-(J-1)/Q\}$ for $Q>0$. The representative variance is generally not the arithmetic mean. It is at least the harmonic mean, with equality only when the variances are equal \citep{Bohning2017}. For finite $J$, the harmonic mean therefore cannot generally replace the representative variance. Also, $I^2$ describes heterogeneity rather than testing for it.

For the independent Gaussian model with fixed hypermean and variance components, the trace of the posterior-mean sensitivity matrix is $\sum_j S_j$. The posterior-mean-deviance measure $p_D$ has the same value for the conditional site-effect likelihood. If the hypermean is instead estimated under a flat prior, the value becomes $1+(J-1)S$ rather than $JS$ for common $S_j=S$ and fixed variances \citep{HodgesSargent2001, Spiegelhalter2002}. These measures describe conditional model complexity, whereas the effective site count $J_{\mathrm{eff}}$ describes how evenly reliability is distributed across sites.

Reliability generalization uses transformations and weights suited to estimated coefficient alpha \citep{FeldtCharter2006, RodriguezMaeda2006, LopezIbanez2024}. Here we give equal weight to sites in a specified trial and study recovery of their effects under three losses. The generalizability index of \citet{Tipton2014} instead concerns how well a sample represents a population of inference. We supplement $I$ with $H$ and $J_{\mathrm{eff}}$ to describe dispersion and concentration within the trial's reliability profile.

\begin{table}[!t]
\caption{Summaries of a site-reliability profile and a related heterogeneity measure.}
\label{tab:crosswalk}
\centering
\begin{tablebody}
\setlength{\tabcolsep}{4pt}
\begin{tabularx}{\linewidth}{@{}>{\raggedright\arraybackslash}p{1.72in}>{\raggedright\arraybackslash}p{1.28in}>{\raggedright\arraybackslash}X@{}}
\toprule
Quantity & Definition & Interpretation \\
\midrule
Reliability $S_j$ & $\sigma^2/(\sigma^2+v_j)$ & Signal as a proportion of total variance; conditional weight on the site's estimate. \\
\addlinespace[5pt]
Pooling weight $\omega_j$ & $1-S_j$ & Conditional weight on the hypermean; general posterior pooling measures may differ. \\
\addlinespace[5pt]
Informativeness $I$ & $\sigma^2/(\sigma^2+\tilde v)$ & Logit-scale center; $\tilde v=\mathrm{GM}(v_j)$. \\
\addlinespace[5pt]
Mean reliability $\bar S$ & $J^{-1}\sum_jS_j$ & Equals $R_b$ for the same inputs; determines the MSEL benchmark with known hyperparameters. \\
\addlinespace[5pt]
Dispersion $H$ & $J^{-1}\sum_j D_\phi(I,S_j)$ & Minimum residual discrepancy under the specified Bernoulli--KL loss. \\
\addlinespace[5pt]
Effective site count $J_{\mathrm{eff}}$ & $(\sum_jS_j)^2/\sum_jS_j^2$ & Concentration expressed as an equivalent site count; does not determine rank certainty or inferential sample size. \\
\addlinespace[5pt]
Variance proportion $R_{\mathrm{typ}}$ & $\sigma^2/(\sigma^2+v_{\mathrm{typ}})$ & Heterogeneity relative to the representative sampling variance in~\eqref{eq:typical-variance}. \\
\bottomrule
\end{tabularx}
\end{tablebody}
\par\vspace{0.4\baselineskip}
\begin{minipage}{\linewidth}\normalsize\onehalfspacing\raggedright
\textit{Note.} The inputs $\sigma^2,v_j$ are fixed and positive, and the equality with $R_b$ requires the same inputs and weights. $R_{\mathrm{typ}}$ differs from the truncated $Q$-based $I^2$ estimate. For $H$, $\phi=\phi_{\mathrm{KL}}$ and $D_\phi(p,q)=\mathrm{KL}\{\Bern(p)\|\Bern(q)\}$, as described in Section~\ref{sec:centroid}. The MSEL benchmark conditions on known hyperparameters.
\end{minipage}
\end{table}

\section{Shrinkage Weights as Reliability Coefficients}\label{sec:reliability}

\subsection{Reliability as a Projection}\label{subsec:reliability-def}

Reliability is commonly expressed as the ratio of signal variance to total variance in multilevel measurement \citep{SnijdersBosker2012}. For square-integrable $X$ and target information $\mathcal G$, \citet{Zumbo2026a} defines the true score as $T=\E[X\mid\mathcal G]$ and reliability as
\begin{equation}\label{eq:rel-def}
\mathrm{Rel}(X) = \frac{\Var(T)}{\Var(X)} = \frac{\Var\!\big(\E[X \mid \mathcal{G}]\big)}{\Var(X)}.
\end{equation}
Since conditional expectation is an orthogonal projection, $T$ is uncorrelated with the error $E=X-T$. When the signal variance is positive, reliability equals the squared cosine of the angle between centered $X$ and $T$ \citep{Zumbo2026a}. This angle is defined for the specified score direction under its joint distribution; it does not describe the norm of the projection operator.

Group-mean reliability uses the mean's sampling variance as error variance \citep{SnijdersBosker2012}. For treatment effects, we take $X_j=\hat\tau_j$, $\mathcal G_j=\sigma(\tau_j)$, and $\E[\hat\tau_j\mid\tau_j]=\tau_j$, with first-stage variance $v_j$. Sites are the objects of measurement and estimated effects are the observed scores. Specifying the objects and sources of error accords with generalizability theory \citep{Cronbach1972, Brennan2001}.

\medskip
\noindent\textbf{Proposition~1 (Shrinkage weights are projection reliabilities).} Under model~\eqref{eq:model}, the reliability of the site estimate $\hat{\tau}_j$ for its true effect $\tau_j$ equals the shrinkage weight:
\[
\mathrm{Rel}(\hat{\tau}_j) = \frac{\Var(\tau_j)}{\Var(\hat{\tau}_j)} = \frac{\sigma^2}{\sigma^2 + v_j} = S_j = \cos^2 \theta_j,
\]
where $\theta_j$ is the angle between the centered $\hat{\tau}_j$ and $\tau_j$ in $L^2$. Equivalently, the conditional pooling weight satisfies $\omega_j = 1 - S_j = \sin^2 \theta_j$. The reliability identity requires only $\E[\hat{\tau}_j \mid \tau_j] = \tau_j$, $\E\{\Var(\hat{\tau}_j \mid \tau_j)\}=v_j$, and finite positive $\sigma^2,v_j$. Normality and independence across sites are unnecessary for this individual ratio. Appendix~\ref{sec:A-reliability} gives the proof and considers the boundary cases $v_j\to0$ and $\sigma^2\to0$.
\medskip

\begin{figure}[!htbp]
\centering
\includegraphics[width=\linewidth]{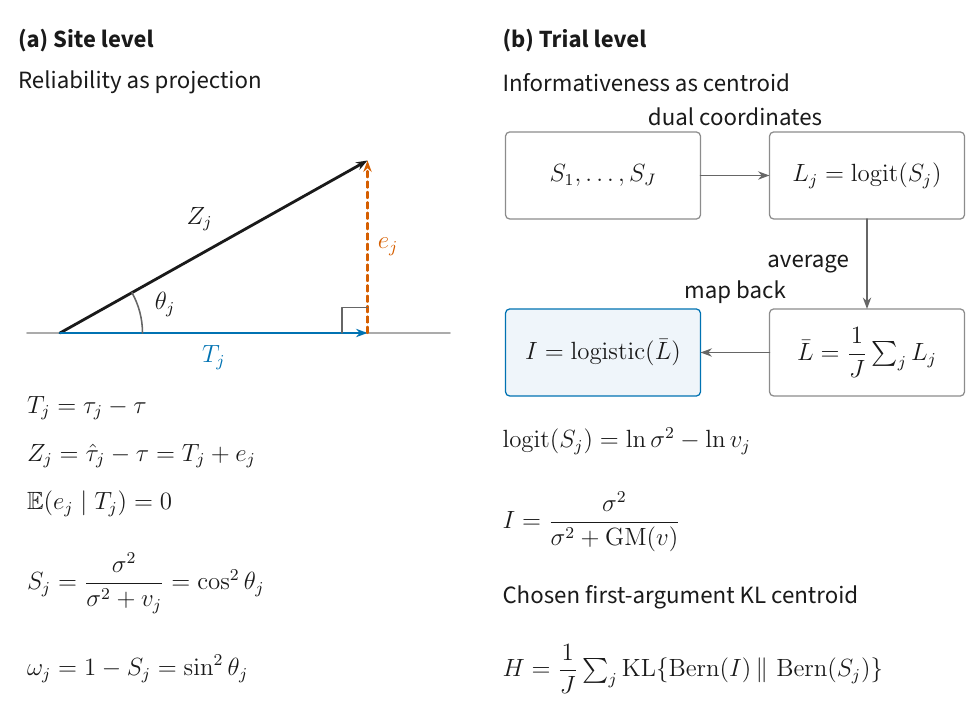}
\caption{Two projection geometries. (a) For centered variables $Z_j=\hat\tau_j-\tau$ and $T_j=\tau_j-\tau$, $Z_j=T_j+e_j$ with $\E(e_j\mid T_j)=0$, giving $S_j=\cos^2\theta_j$ (Proposition~1). (b) Informativeness is obtained by averaging reliability logits across sites and is the left Bernoulli--KL centroid (Theorem~1). The dispersion $H$ measures the minimum discrepancy from a constant profile under this loss (Section~\ref{sec:dispersion}).}
\label{fig:geometry}
\end{figure}

Proposition~1 uses the conditional expectation $\E[\hat\tau_j\mid\tau_j]$. The posterior mean instead uses $\E[\tau_j\mid\hat\tau_j]$, reversing the conditioning. These moment conditions do not imply the normal--normal posterior rule~\eqref{eq:posterior} under a non-Gaussian model.

In Figure~\ref{fig:geometry}(a), the centered estimate is the hypotenuse of a right triangle and its projection is the centered true effect. Squared lengths represent variances, so reliability is the fraction of the total squared length retained by the projection. A smaller angle means less sampling error relative to signal. The corresponding odds equal the classical signal-to-noise ratio \citep{CronbachGleser1964}:
\[
\frac{S_j}{1-S_j}=\frac{\sigma^2}{v_j}.
\]
Reliability logits are therefore log signal-to-noise ratios. Aggregation across sites requires a further choice of divergence and averaging rule.

\section{The Informativeness Index as a Kullback--Leibler Centroid}\label{sec:centroid}

\subsection{Averaging Reliabilities on Different Scales}\label{subsec:aggregation}

We seek a single reliability to represent the sites. What counts as the closest common value depends on how we measure discrepancy. The arithmetic mean $\bar S=J^{-1}\sum_jS_j$ averages reliabilities on their original scale; informativeness uses the geometric mean of sampling variances. We ask which discrepancy makes informativeness the closest common reliability, or \emph{centroid}. For a continuously differentiable, strictly convex potential $\phi$ on $(0,1)$, the Bregman divergence is
\begin{equation}\label{eq:bregman}
D_\phi(x,y)=\phi(x)-\phi(y)-\phi'(y)(x-y)\geq0
\end{equation}
\citep{Bregman1967,Banerjee2005}. Geometrically, this is the vertical gap between the convex curve and its tangent at the second argument, evaluated at the first argument. Reversing the arguments changes the tangent, so the divergence is generally asymmetric. Minimizing $J^{-1}\sum_jD_\phi(p,S_j)$ with respect to the first argument gives the \emph{left centroid},
\[
I_\phi=(\phi')^{-1}\!\left\{J^{-1}\sum_j\phi'(S_j)\right\},
\]
whereas minimizing $J^{-1}\sum_jD_\phi(S_j,p)$ with respect to the second argument gives the \emph{right centroid} $\bar S$ \citep{NielsenNock2009,NielsenNock2020}. Both minimizations have unique interior solutions in this scalar setting (Appendix~\ref{sec:B-centroid}). To obtain the left centroid, we transform the reliabilities by the derivative of the potential, take their arithmetic mean, and transform back. These transformed values are called dual coordinates.

We use negative Bernoulli entropy as the convex potential,
\[
\phi_{\mathrm{KL}}(p)=p\ln p+(1-p)\ln(1-p),
\]
whose derivative is $\operatorname{logit}(p)$ and whose divergence is $\mathrm{KL}\{\Bern(p)\|\Bern(q)\}$ \citep{Amari2016}. We treat reliabilities as Bernoulli parameters only to define the profile loss, leaving the sampling model unchanged. We obtain the left centroid by averaging the reliability logits and applying the inverse logit, as in Figure~\ref{fig:geometry}(b). The result equals the informativeness index.

\medskip
\noindent\textbf{Theorem~1 (Informativeness as a centroid).} For a finite uniformly weighted profile with common $\sigma^2>0$ and fixed $v_j>0$, let $S_j=\sigma^2/(\sigma^2+v_j)$. The informativeness~\eqref{eq:informativeness} of \citet{LeeEtAl2025} is the unique left Bernoulli--KL centroid:
\[
\begin{gathered}
I=\operatorname*{arg\,min}_{p\in(0,1)}\frac1J\sum_{j=1}^J
\mathrm{KL}\{\Bern(p)\|\Bern(S_j)\},\\
\operatorname{logit}(I)=\frac1J\sum_j\operatorname{logit}(S_j).
\end{gathered}
\]
Indeed, $\operatorname{logit}(S_j)=\ln\sigma^2-\ln v_j$, so averaging the dual coordinates gives $I=\sigma^2/(\sigma^2+\tilde v)$. The full proof is in Appendix~\ref{subsec:B-proof}.
\medskip

Consider the class of fixed continuously differentiable, strictly convex scalar generators. Requiring the first-argument centroid to equal the logit mean for \emph{every interior profile consisting of two equally weighted points} identifies $\phi_{\mathrm{KL}}$ up to positive scale and affine terms (Appendix~\ref{sec:B-centroid}, Lemma~\ref{lem:B-logit-characterization}). Boundedness of the reliabilities alone, or agreement for one observed profile, is insufficient. The requirement neither fixes the divergence units nor establishes optimality under an empirical estimation loss.

Table~\ref{tab:centroids} compares three choices of potential. The harmonic centroid is no greater than either $I$ or $\bar S$, whereas $I$ and $\bar S$ have no universal ordering. We interpret their difference in light of profile shape and the loss of interest. Under the KL loss, the residual discrepancy measures dispersion and relates this difference to estimation risk under specified conditions.

\begin{table}[!t]
\caption{First-argument Bregman centroids of an interior reliability profile.}
\label{tab:centroids}
\centering
\begin{tablebody}
\setlength{\tabcolsep}{4pt}
\begin{tabularx}{\linewidth}{@{}>{\raggedright\arraybackslash}p{2.00in}>{\raggedright\arraybackslash}p{1.12in}>{\raggedright\arraybackslash}X@{}}
\toprule
Potential $\phi(x)$ & Dual coordinate $\phi'(x)$ & Left centroid $I_\phi$ \\
\midrule
$\tfrac12x^2$\newline Squared Euclidean & $x$ & Arithmetic mean $\bar S=J^{-1}\sum_jS_j$ \\
\addlinespace[5pt]
$x\ln x+(1-x)\ln(1-x)$\newline Negative Bernoulli entropy & $\operatorname{logit}(x)$ & $\operatorname{logistic}\!\left\{J^{-1}\sum_j\operatorname{logit}(S_j)\right\}=I$ \\
\addlinespace[5pt]
$-\ln x$\newline Burg potential & $-1/x$ & Harmonic mean $\operatorname{HM}(S)=\left(J^{-1}\sum_jS_j^{-1}\right)^{-1}$ \\
\bottomrule
\end{tabularx}
\end{tablebody}
\par\vspace{0.4\baselineskip}
\begin{minipage}{\linewidth}\normalsize\onehalfspacing\raggedright
\textit{Note.} Each centroid minimizes $J^{-1}\sum_jD_\phi(p,S_j)$ with respect to its first argument. Minimizing with respect to the second argument gives $\bar S$ for all three potentials \citep{NielsenNock2009,NielsenNock2020}. For an interior profile, $\operatorname{HM}(S)\leq I$ and $\operatorname{HM}(S)\leq\bar S$. All three means coincide if and only if the profile is constant. There is no universal ordering of $I$ and $\bar S$, and equality of these two means does not imply $H=0$. For example, the profile $(.2,.8)$ gives $I=\bar S=.5$ and $H=\ln(5/4)>0$. Further details are given in Appendices~\ref{sec:B-centroid} and~\ref{sec:C-dispersion}.
\end{minipage}
\end{table}

\section{Dispersion and Concentration Diagnostics}\label{sec:dispersion}

\subsection{Dispersion: What a Constant Profile Leaves Out}\label{subsec:pythagoras}

\begin{figure}[!htbp]
\centering
\includegraphics[width=\linewidth]{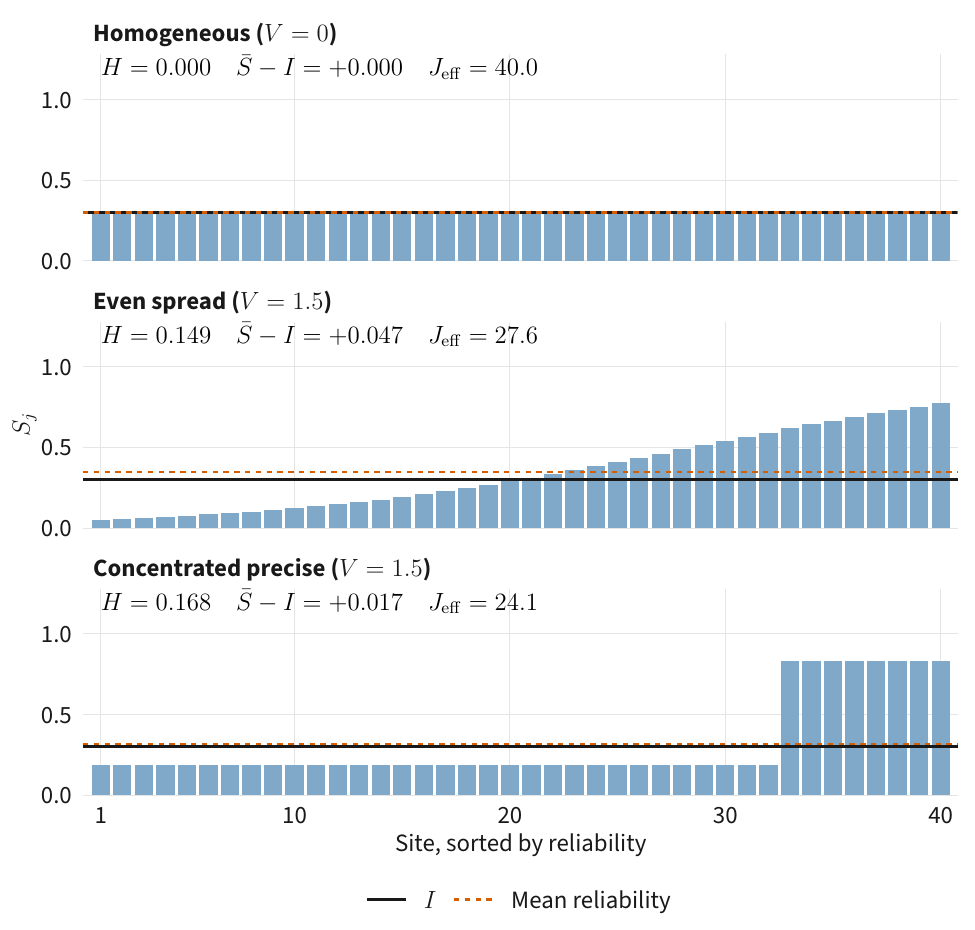}
\caption{Three reliability profiles with identical informativeness. Each panel shows sorted site reliabilities for $J=40$ and $I=.30$. The solid line marks $I$ and the dashed line marks $\bar S$. The difference between them determines how MSEL compares with that of a homogeneous profile under the conditions of Proposition~2. The annotations give logit variance $V$, Bernoulli--KL dispersion $H$, the difference $\bar S-I$, and effective site count $J_{\mathrm{eff}}$.}
\label{fig:profiles}
\end{figure}

A constant reliability summarizes the profile by replacing $(S_1,\ldots,S_J)$ with $(p,\ldots,p)$. Figure~\ref{fig:profiles} shows three profiles with informativeness $I=.30$ and different spreads and concentrations. Using the Bernoulli--KL loss of Section~\ref{sec:centroid}, we define dispersion as
\begin{equation}\label{eq:dispersion}
H=\frac1J\sum_{j=1}^J\mathrm{KL}\{\Bern(I)\|\Bern(S_j)\}\ge0.
\end{equation}
This is the minimum average discrepancy over constant profiles for the specified loss and equals zero if and only if all $S_j$ are equal.

\medskip
\noindent\textbf{Theorem~2 (Pythagorean decomposition for constant profiles).} Let $\phi$ be continuously differentiable and strictly convex on an open interval containing $S_1,\ldots,S_J$, and let $I_\phi$ be their first-argument centroid. For every candidate $p$ in that interval,
\[
\frac1J\sum_jD_\phi(p,S_j)=D_\phi(p,I_\phi)+H_\phi,
\qquad H_\phi=\frac1J\sum_jD_\phi(I_\phi,S_j).
\]
Conversely, if this identity holds for every $p$ with a center $c$ and a constant remainder, then $c=I_\phi$ and the remainder is $H_\phi$. Under $\phi_{\mathrm{KL}}$, the remainder is $H$ in~\eqref{eq:dispersion}.
\medskip

This resembles the familiar decomposition of a sum of squared deviations about a candidate value into variation about the mean and a penalty for moving away from the mean. Under the KL loss, $H$ is the discrepancy remaining at the best constant profile; choosing another constant adds to it. The divergences need not be squared distances. The cross term vanishes because the gradient at the centroid equals the mean gradient (Appendix~\ref{sec:C-dispersion}). Thus $H$ describes what the constant profile leaves out under the stated loss and weights. It is a Jensen gap in dual coordinates and generally differs from the gap obtained by reversing the divergence \citep{NielsenNock2020}.

\subsection{Dispersion and Benchmark Estimation Risk}\label{subsec:risk}

Increasing the common signal variance shifts all log signal-to-noise ratios by the same amount, preserving the gaps between them. Along this path, the difference between the arithmetic mean $\bar S$ and the logit mean $I$ gives the rate of change of $H$ with the mean logit. Write $m=\operatorname{logit}(I)$ and $\Delta_j=\operatorname{logit}(S_j)-m$, with $J^{-1}\sum_j\Delta_j=0$, and define $V=J^{-1}\sum_j\Delta_j^2$ and $\rho_3=J^{-1}\sum_j|\Delta_j|^3$. Since $\operatorname{logit}(S_j)=\ln\sigma^2-\ln v_j$, $V$ depends only on the $v_j$ and remains fixed as $\sigma^2$ varies. We condition on the supplied variances, even when the $v_j$ are estimated.

\medskip
\clearpage
\noindent\textbf{Proposition~2 (Dispersion and estimation risk).} Hold the centered logits $\Delta_j$ fixed. With $A(t)=\ln(1+e^t)$:
\begin{align*}
\text{(i)}\quad H(m;\Delta)&=\frac1J\sum_j A(m+\Delta_j)-A(m),\\
\frac{\partial H}{\partial m}&=\bar S-I,
\qquad \frac{\partial H}{\partial I}=\frac{\bar S-I}{I(1-I)}.
\end{align*}
The second-order expansions have explicit remainders:
\begin{align*}
\text{(ii)}\quad H&=\tfrac12 I(1-I)V+R_H,
& |R_H|&\le\frac{\sqrt3}{108}\rho_3,\\
\bar S-I&=\tfrac12 I(1-I)(1-2I)V+R_S,
& |R_S|&\le\frac{\rho_3}{48}.
\end{align*}
Suppose that the common mean $E(\tau_j)=\tau$ and $\sigma^2,v_j$ are known and that the conditional moment assumptions of Proposition~1 hold. The minimum integrated MSEL among predictors affine in each site's own estimate is
\[
\text{(iii)}\quad \mathcal R_{\mathrm{aff}}
=\sigma^2(1-\bar S)
=\sigma^2\left\{1-I-\frac{\partial H}{\partial m}\right\}.
\]
In the independent Gaussian model~\eqref{eq:model}, the same predictor is the posterior mean conditional on known hyperparameters, and this risk also equals the average posterior variance. Appendix~\ref{sec:C-dispersion} gives the proof and the risk approximation with its remainder scaled by the variance.
\medskip

Part (i) holds with centered logits fixed; it does not describe arbitrary changes in site size or the risk of a fitted model. In part (ii), $\bar S-I$ has the sign of $1-2I$ provided that the absolute leading term exceeds the remainder bound $\rho_3/48$. Near $I=1/2$, or when the profile is strongly skewed, the remainder can determine the sign.

For example, the $J=25$ profile in Appendix~\ref{subsec:C-remarks} has $I=.30$ and $V=2$, yet $\bar S=.2732468<I$. Reversing its centered logits at $I=.70$ gives $\bar S=.7267532>I$. Compared with a homogeneous profile having the same $I$ and $\sigma^2$, the benchmark risk is lower if and only if $\bar S>I$. There is thus no universal ordering of the two means, and the difference between them need not increase with $H$.

For independent Gaussian sites, part (iii) gives the Bayes risk with known hyperparameters \citep{Morris1983}. Under conditional moment assumptions alone, optimality holds within the class of affine predictors based on each site's own estimate; it need not extend to predictors using the full vector. This risk integrates over site effects with hyperparameters fixed; it is neither conditional on realized effects nor the risk after fitting hyperparameters. At a fixed geometric mean, unequal sampling variances have greater total precision than equal variances. Resources therefore differ, and we examine fixed enrollment separately in Section~\ref{subsec:sim-fixedN}. The role of $\bar S$ is specific to this MSEL benchmark; ranks and EDFs require separate comparisons.

\subsection{An Effective Site Count for the Reliability Profile}\label{subsec:jeff}

We express how evenly reliability is distributed across sites as an effective site count. Normalize each site's reliability as a share of the total, $q_j=S_j/\sum_kS_k$, and define
\begin{equation}\label{eq:jeff}
J_{\mathrm{eff}}=\frac{1}{\sum_jq_j^2}
=\frac{\big(\sum_jS_j\big)^2}{\sum_jS_j^2}.
\end{equation}
A reference profile of $k$ equally reliable sites has shares $1/k$ and an effective count of $k$. The count therefore expresses the concentration of normalized reliability as an equivalent number of equally contributing sites. Larger values indicate less concentration.

For an interior profile with $J>1$, $1<J_{\mathrm{eff}}\le J$. The upper bound $J$ is attained if and only if the profile is constant, and the lower endpoint is approached as one normalized weight tends to one. For $J=1$, the value is one.

Multiplying all reliabilities by a common positive constant leaves this count unchanged, provided the rescaled values remain interior. For example, 40 sites each with reliability .01 and 40 sites each with reliability .90 both give $J_{\mathrm{eff}}=40$. We therefore interpret the count alongside $I$ and $\bar S$ to distinguish relative contributions from the overall level of reliability. It does not measure inferential sample size, model complexity, rank recovery, or sensitivity of distribution estimation. A small value warrants inspecting the sites with the largest reliability shares.

\subsection{Estimating the Diagnostics}\label{subsec:estimation}

We can average a diagnostic over the posterior of $\sigma^2$, or first average the variance and then calculate the diagnostic. The two operations need not agree. For $I$, the second calculation gives a value at least as large because informativeness increases at a decreasing rate.

\medskip
\noindent\textbf{Proposition~3 (Posterior Jensen comparison).} Fix $0<v_j<\infty$ and write $a=\sigma^2\ge0$ and $c=\tilde v>0$. The function $I(a)=a/(a+c)$ is strictly concave on $[0,\infty)$. For a proper posterior with $\E(a\mid\text{data})<\infty$,
\[
\E\{I(a)\mid\text{data}\}\le I\{\E(a\mid\text{data})\},
\]
with equality if and only if the posterior of $a$ is degenerate. The same comparison holds for each $S_j(a)=a/(a+v_j)$. Appendix~\ref{sec:D-estimation} gives the proof and a variance-based approximation with an explicit remainder condition.
\medskip

Proposition~3 compares two summaries of the same posterior; it has no implication for their frequentist biases. We report posterior means of the diagnostics and central credible intervals. The proposition neither orders the summaries of $H$ and $J_{\mathrm{eff}}$ nor guarantees repeated-sampling coverage.

\begin{figure}[!htbp]
\centering
\includegraphics[width=\linewidth]{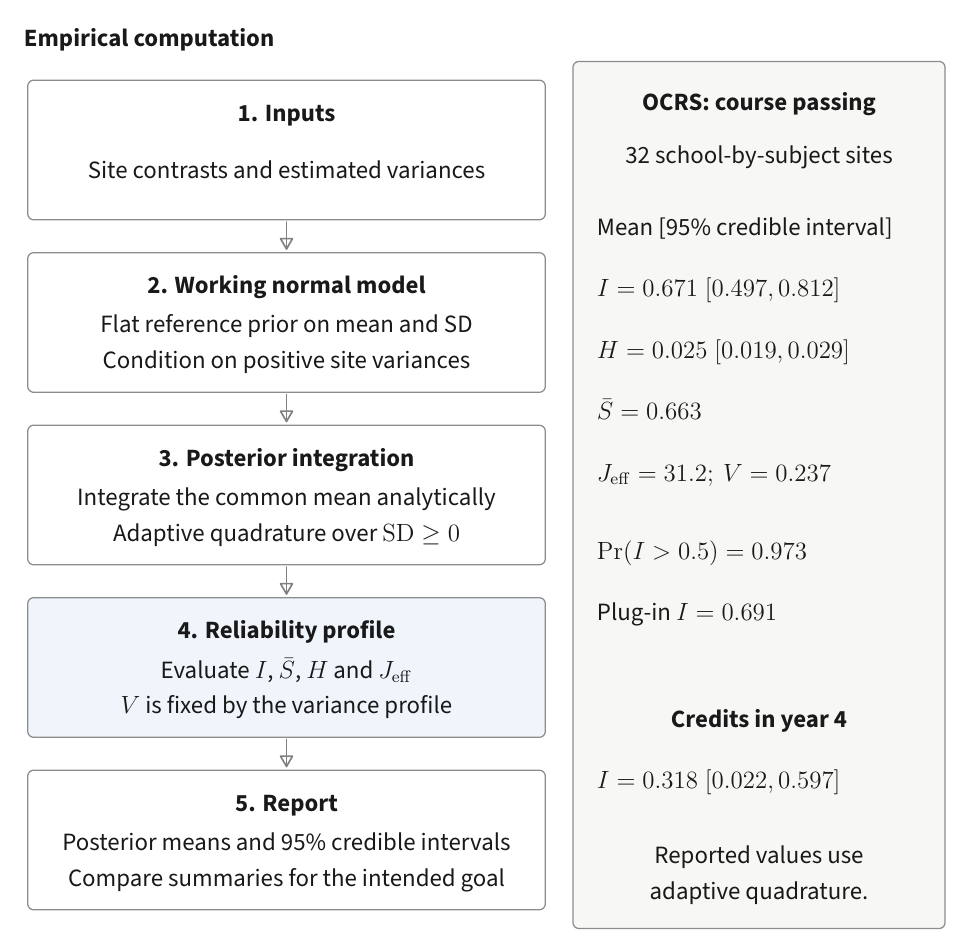}
\caption{Computation of posterior reliability diagnostics. The left column shows how the diagnostics are obtained from first-stage inputs under the working model. The right column illustrates the procedure for the OCRS outcomes ($J=32$; Table~\ref{tab:ocrs}). Appendix~\ref{sec:D-estimation} describes the adaptive integration used for the application and the grid procedure used for the simulations.}
\label{fig:workflow}
\end{figure}

For the independent normal--normal reference model, we use $p(\tau,\sigma)\propto1$ on $\mathbb R\times[0,\infty)$. The prior is improper and flat in the mean and heterogeneity \emph{SD}. A flat prior depends on the chosen scale, so this parameterization matters \citep{Gelman2006}. For finite data and fixed positive $v_j$, the posterior is proper if and only if $J>2$, and $\E(\sigma^2\mid\text{data})$ is finite if and only if $J>4$ (Appendix~\ref{subsec:D-prior}). All reported fits satisfy these conditions. The marginal posterior of $\sigma$ is one-dimensional, allowing direct integration \citep{Rubin1981,Gelman2013}. We use adaptive quadrature for the application and for posterior means in the bootstrap and subset analyses. For the simulations, we use a $250$-point grid and $2{,}000$ draws per fit. Appendix~\ref{subsec:D-algorithm} describes both algorithms and their checks, and Figure~\ref{fig:workflow} illustrates the computation.

Integrating over $\sigma^2$ accounts for heterogeneity uncertainty, conditional on the supplied first-stage variances. Uncertainty in these inputs remains unmodeled. The bootstrap in Appendix~\ref{sec:G-ocrs} assesses the spread of refitted summaries under the specified resampling scheme.

\section{Design Implications}\label{sec:design}

We compare enrollment allocations by informativeness, fixing the number of sites $J$, total enrollment $N$, and a common positive signal variance $\sigma^2$. Suppose that $v_j=\kappa/n_j$, with the same $\kappa>0$ at every site and equal enrollment costs. This special case of~\eqref{eq:neyman} need not hold when treatment proportions or outcome variances differ across sites.

\begin{samepage}
\medskip
\noindent\textbf{Proposition~4 (Balanced allocation maximizes informativeness).} Among positive continuous allocations with $\sum_jn_j=N$ and fixed $J$, $I$ is uniquely maximized by $n_j=N/J$ for all $j$. The maximum is $I=\sigma^2/(\sigma^2+\kappa J/N)$. Under positive integer constraints, a maximizing allocation makes site sizes as equal as possible. Under the same conditions, balance also maximizes $\bar S$ and minimizes the risk $\sigma^2(1-\bar S)$ among predictors that are affine in each site's own estimate, with the inputs treated as known. The proof is given in Appendix~\ref{sec:E-design}.
\medskip
\end{samepage}

For $I$, the result follows from the arithmetic--geometric mean inequality, and for $\bar S$ it follows from concavity. These results do not establish optimal allocations for ranks, EDF estimation, or fitted hierarchical procedures. Other design objectives concern population-mean effects, site sampling, and the number of sites \citep{BloomSpybrook2017,Miratrix2021}, and can favor different allocations. Increasing $J$ at fixed $N$ reduces $I$ in a balanced design but can improve precision for a population mean (Appendix~\ref{sec:E-design}).

\begin{figure}[!htbp]
\centering
\includegraphics[width=\linewidth]{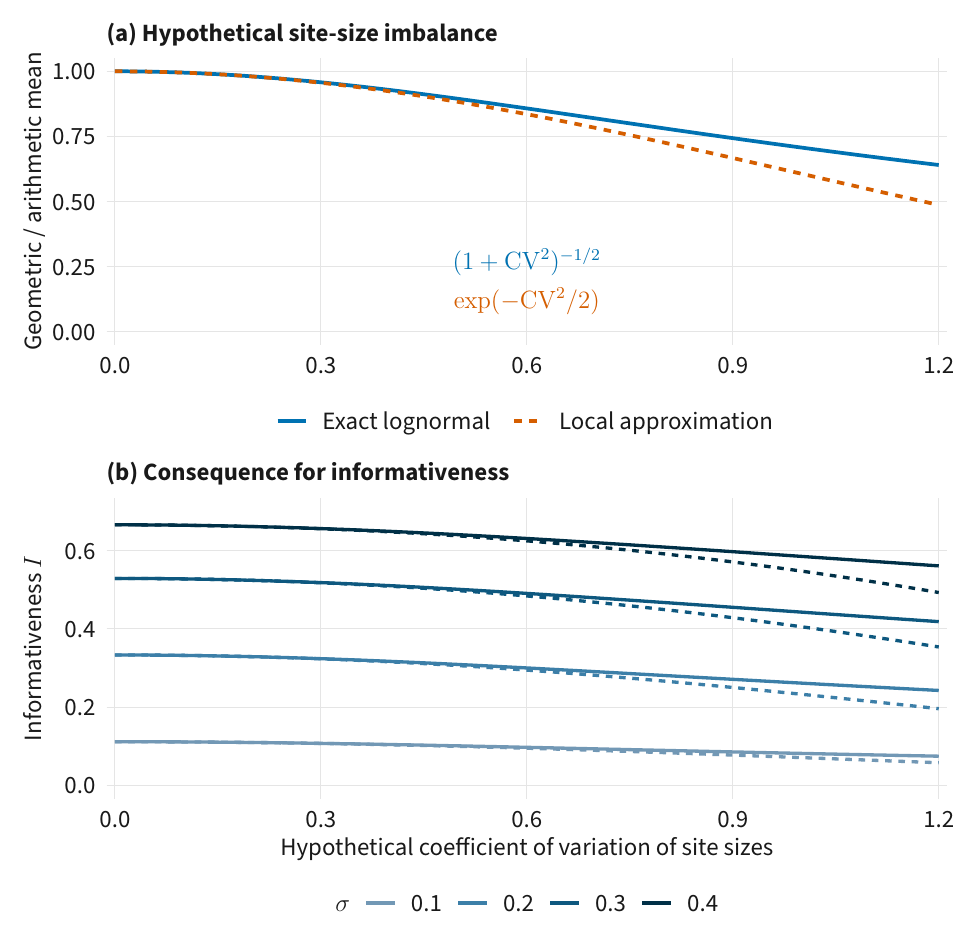}
\caption{Hypothetical imbalance in site sizes. (a) The geometric-to-arithmetic mean ratio for a lognormal population (solid) and its local approximation (dashed). (b) The corresponding informativeness for $v=4/n$, mean size $50$, and the indicated signal SDs. The exact ratio is $(1+\mathrm{CV}^2)^{-1/2}$ and the approximation is $\exp(-\mathrm{CV}^2/2)$. These population calculations do not give identities for finite profiles and are not calibrated to education trials.}
\label{fig:design}
\end{figure}

For a finite allocation, the geometric mean $\tilde n=\mathrm{GM}(n_j)$ is an \emph{equivalent balanced site size}: a hypothetical balanced trial with this size at every site has the same $I$. The comparison holds the site count, signal variance, and common variance law fixed and allows continuous sizes. The equivalent size is no greater than the actual average site size; it describes informativeness, not efficiency for every estimation goal. Write $n_j=\bar n(1+u_j)$, with $\sum_j u_j=0$. For small relative deviations, the local approximation is $\tilde n/\bar n\approx\exp(-\mathrm{CV}_J^2/2)$, where $\mathrm{CV}_J^2=J^{-1}\sum_ju_j^2$. Outside this local regime, CV alone does not determine the ratio. For a lognormal \emph{population} of sizes, the geometric-to-arithmetic mean ratio is exactly $(1+\mathrm{CV}^2)^{-1/2}$. This population identity does not hold for arbitrary finite profiles. Figure~\ref{fig:design} compares the exact calculation with the local approximation in hypothetical scenarios that are not calibrated to typical educational trials. We examine fixed enrollment by simulation in Section~\ref{subsec:sim-fixedN}.

\section{Simulation Study}\label{sec:simulation}

\subsection{Questions and Design}\label{subsec:sim-design}

We first ask whether trials with the same informativeness can differ in estimation accuracy. We then examine which profile summaries describe these differences and how estimator performance depends on the inferential goal. A separate experiment assesses unequal site enrollment when total enrollment is fixed.

To compare profiles at the same $I$, we hold the between-site variance and the geometric mean $\mathrm{GM}(v_j)$ fixed and vary dispersion and shape. These comparisons do not hold resources fixed: at $V=2$, total precision $\sum_j1/v_j$ is $1.6$ to $3.8$ times the value for a homogeneous profile. In the second experiment, we fix total enrollment $N=J\bar n$ and set $v_j=4/n_j$. We obtain hypothetical continuous site sizes by clipping and rescaling Gamma profiles. Their construction uses the parent Gamma coefficient of variation (CV), which differs from the realized finite-site CV and is not calibrated to educational trials.

We compare the unpooled estimates, posterior mean (PM), constrained Bayes (CB), and triple-goal (GR) estimates for three targets: site effects (RMSE), ranks (MSELP), and the finite effect distribution (ISEL). Table~\ref{tab:simdesign} gives the settings. We generated Gaussian, mixture, and asymmetric Laplace effects with \textsf{multisiteDGP} \citep{multisiteDGP}, using $500$ replications per cell across $702+108$ cells. In the fixed-$I$ experiment, profiles within each $(J,I,G)$ family, where $G$ denotes the generating effect distribution, share the same latent effects and standardized sampling errors. We can therefore compare their losses in pairs.

The fitted shrinkage estimators use the normal working hierarchy with a flat reference prior on the mean and SD, evaluated on a $250$-point grid with $2{,}000$ posterior draws. We use conditional moments (Rao--Blackwellization) to preserve the exact equality of ranks for homogeneous profiles. In six selected Gaussian cases, we compared the limiting grid and jitter distribution with quadrature. For PM and CB, using quadrature changed RMSE by at most $0.244\%$ and ISEL by at most $1.185\%$. These checks do not cover GR, the full grid, or variation from finite draws; the results compare the procedures as implemented (Appendices~\ref{sec:D-estimation} and~\ref{sec:F-simulation}).

Both experiments include the linear benchmark with known $(\tau,\sigma^2)$, whose integrated MSEL is $\sigma^2(1-\bar S)$ by Proposition~2. It is the posterior mean for independent Gaussian site effects; for the other generating distributions, it remains the affine benchmark based on a site's own estimate. The fixed-$I$ experiment also includes a semi-oracle with known $\sigma^2$ and an estimated hypermean.

\begin{table}[!t]
\caption{Simulation settings for profiles with fixed $(I,V)$ and profiles with fixed total enrollment.}
\label{tab:simdesign}
\centering
\begin{tablebody}
\setlength{\tabcolsep}{4pt}
\begin{tabularx}{\linewidth}{@{}>{\raggedright\arraybackslash}p{1.72in}>{\raggedright\arraybackslash}X@{}}
\toprule
Factor & Levels \\
\midrule
\multicolumn{2}{@{}l}{\emph{Fixed informativeness: 702 cells}} \\
\addlinespace[5pt]
Sites $J$ & $25$, $50$, $100$ \\
\addlinespace[5pt]
Informativeness $I$ & $.05$, $.15$, $.30$, $.45$, $.60$, $.75$ \\
\addlinespace[5pt]
Logit dispersion $V$ & $0$, $.5$, $1$, $2$ \\
\addlinespace[5pt]
Shape for $V>0$ & Even; bimodal; concentrated precise; concentrated noisy. Shapes coincide at $V=0$. \\
\midrule
\multicolumn{2}{@{}l}{\emph{Fixed enrollment: 108 cells}} \\
\addlinespace[5pt]
Sites and mean size & $J\in\{25,50,100\}$, $\bar n\in\{20,50,100\}$; $N=J\bar n$, $v_j=4/n_j$ \\
\addlinespace[5pt]
Parent Gamma CV & $0$, $.25$, $.50$, $.75$; continuous sizes obtained by clipping and rescaling twice (Appendix~\ref{subsec:F-fixedN}) \\
\midrule
\multicolumn{2}{@{}l}{\emph{Common settings}} \\
\addlinespace[5pt]
True $G$ & Gaussian; two-component Gaussian mixture; asymmetric Laplace \\
\addlinespace[5pt]
SD and replications & $\sigma=.20$; 500 replications per cell \\
\addlinespace[5pt]
Procedures & ML, PM, CB, GR; known-$(\tau,\sigma^2)$ linear benchmark in both experiments; known-$\sigma^2$ semi-oracle for fixed informativeness \\
\addlinespace[5pt]
Losses & RMSE, MSELP, ISEL for the finite effects \\
\bottomrule
\end{tabularx}
\end{tablebody}
\par\vspace{0.4\baselineskip}
\begin{minipage}{\linewidth}\normalsize\onehalfspacing\raggedright
\textit{Note.} The stress profiles attain the specified $(I,V)$ exactly. Increasing dispersion at fixed geometric mean also increases total precision, so the two sets of simulations address different questions. Profiles within families use paired data draws. Fits use a normal model, a flat reference prior on the mean and SD, 250 nodes, $M=2{,}000$, and the marginal-variance approximation for CB. Data were generated with \textsf{multisiteDGP} 0.1.0 \citep{multisiteDGP}. Construction and aggregation are described in Appendix~\ref{sec:F-simulation}.
\end{minipage}
\end{table}

\subsection{Different Performance at the Same Informativeness}\label{subsec:sim-landscape}

\begin{figure}[!htbp]
\centering
\includegraphics[width=\linewidth]{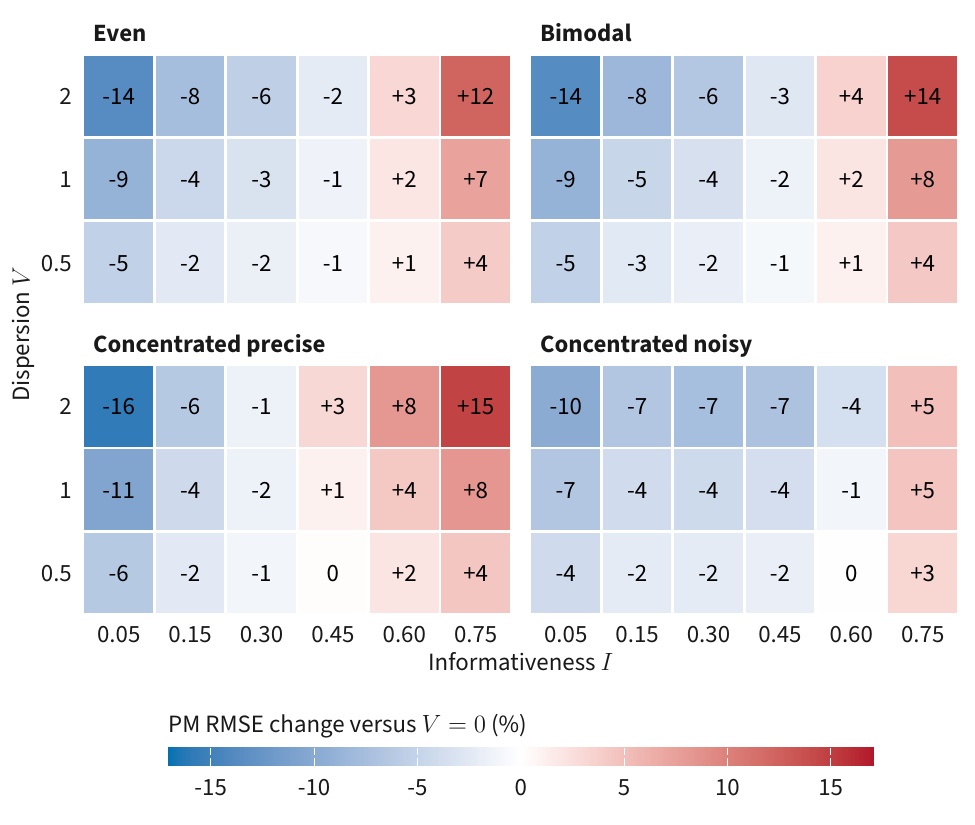}
\caption{Percentage changes in PM RMSE calculated from geometric paired ratios relative to the homogeneous profile at the same $(I,J,G)$, averaged over the fixed $J\times G$ grid. Blue indicates lower loss and red higher loss. The crossing depends on profile shape. The local result for affine MSEL does not determine every comparison of fitted RMSE.}
\label{fig:landscape}
\end{figure}

Profiles with the same $I$ can give different RMSEs, and greater dispersion can either lower or raise loss (Figure~\ref{fig:landscape}). We compare each dispersed profile with the homogeneous profile at the same $(J,I,G)$. A ratio below one indicates lower RMSE; a ratio above one indicates higher RMSE.

For even profiles at $V=2$, the PM ratios range from $.86$ at $I=.05$ to $1.12$ at $I=.75$. Both even and bimodal profiles cross from lower to higher RMSE between $I=.45$ and $.60$. Shape also matters: at $I=.45$ and $V=2$, RMSE increases by $3.1\%$ for concentrated-precise profiles but decreases by $7.0\%$ for concentrated-noisy profiles (Monte Carlo standard error, MCSE, about $0.1$ percentage point in each case). The aggregate pattern resembles the comparison of affine risks with known inputs, although the direction for fitted RMSE differs in $9$ of the $648$ nonconstant family-by-profile cells. Thus neither $I$ nor $I$ together with $V$ determines the direction of the difference.

We summarize these comparisons with geometric paired ratios. We first average paired log-loss ratios within each $(J,I,G)$ family, then average across the nine $J\times G$ families at a given $I$ and exponentiate. This generally differs from a ratio of arithmetic mean losses. The MCSEs account for paired observations and seeds shared across posterior fits (Appendix~\ref{sec:F-simulation}).

Estimating the parameters also changes the comparison. At $I=.05$, the ratios for even profiles at $V=2$ are $.976$ for the linear estimator with known inputs, $.919$ for the semi-oracle, and $.863$ for PM. Each ratio compares dispersed and homogeneous profiles within the same procedure. The ratios do not compare absolute losses across procedures or isolate the causal contribution of each estimation step.

\subsection{What Profile Summaries Add to Informativeness}\label{subsec:sim-metamodel}

\begin{table}[!t]
\caption{Variation in cell-mean log losses explained by reliability profiles.}
\label{tab:metamodel}
\centering
\begin{tablebody}
\begin{tabularx}{\linewidth}{@{}l*{4}{>{\raggedleft\arraybackslash}X}@{}}
\toprule
\multicolumn{5}{@{}l}{\emph{A. In-sample partial $R^2$ for added profile factors}} \\
\addlinespace[5pt]
Loss & PM & CB & GR & ML \\
\midrule
RMSE & .16 & .15 & .15 & $>$.99 \\
\addlinespace[5pt]
MSELP & .32 & .32 & .32 & .59 \\
\addlinespace[5pt]
ISEL & .22 & .36 & .52 & .76 \\
\addlinespace[10pt]
\multicolumn{5}{@{}l}{\emph{B. Five-fold CV $R^2$ with cells held out: selected comparisons}} \\
\addlinespace[5pt]
Loss & \multicolumn{1}{l}{Estimator} & $+H$ & $+V$ & $+\bar S$ \\
\midrule
RMSE & \multicolumn{1}{l}{PM} & .98964 & .99155 & .99843 \\
\addlinespace[5pt]
MSELP & \multicolumn{1}{l}{GR} & .99298 & .99470 & .99919 \\
\addlinespace[5pt]
ISEL & \multicolumn{1}{l}{GR} & .99586 & .99572 & .99518 \\
\bottomrule
\end{tabularx}
\end{tablebody}
\par\vspace{.4\baselineskip}
\begin{minipage}{\linewidth}\normalsize\onehalfspacing\raggedright
\textit{Note.} $I$, $H$, $V$, and $\bar S$ are calculated from the true generating profiles. Panel A reports $(R^2_\mathrm{full}-R^2_\mathrm{base})/(1-R^2_\mathrm{base})$. The base model is saturated in $(I,J,G)$, and the full model adds crossed $V$-by-shape factors. In Panel B, each summary $X$ enters as $X+X\times\mathrm{factor}(I)$, using the same model structure and cell folds. All nine comparisons are given in Table~\ref{tab:F-fair}. PM denotes the posterior mean, CB the implemented constrained-Bayes approximation, GR the triple-goal estimator, and ML the unpooled estimates. The results refer to the simulation grid.
\end{minipage}
\end{table}

\begin{figure}[!htbp]
\centering
\includegraphics[width=\linewidth]{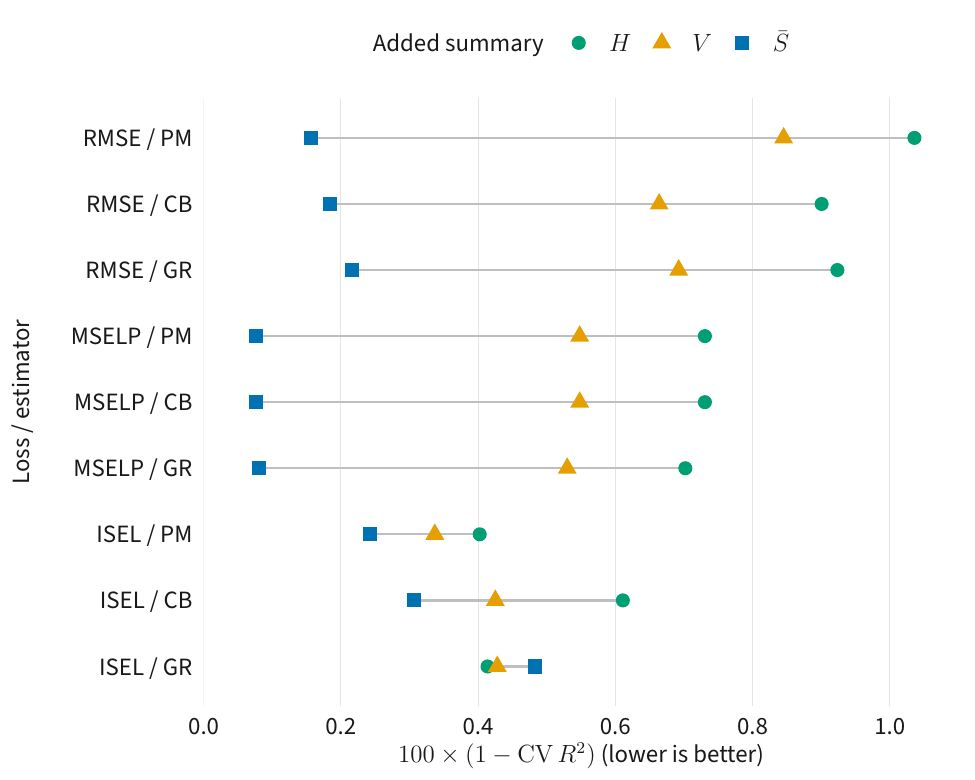}
\caption{Comparison of single summaries of the true generating profiles using models with the same structure for nine combinations of inferential goal and estimator. The axis gives $100(1-\mathrm{CV}\ R^2)$, so the leftmost marker has the largest CV $R^2$. Five-fold cross-validation holds out cells within the simulation grid. $\bar S$ has the largest value in eight comparisons and $H$ for GR's ISEL. Uncertainty in these orderings has not been assessed.}
\label{fig:metamodel}
\end{figure}

The metamodels use the generating values of $I$, $H$, $V$, and $\bar S$, rather than estimates from the simulated data. We regress cell-mean log losses on a base model saturated in $(I,J,G)$ and add crossed dispersion and shape factors (Table~\ref{tab:metamodel}, Panel A). These factors account for $15$--$52\%$ of the remaining sum of squares for the shrinkage estimators and up to $99.9\%$ for the unpooled estimates. These partial $R^2$ values measure improvement in fit within the simulation grid. Additional interactions could change the results, and the remaining variation cannot be attributed solely to Monte Carlo error.

We compare $H$, $V$, and $\bar S$ as predictors of losses in held-out cells, using models with the same structure and degrees of freedom (Panel B). Five-fold cross-validation retains other profiles from the same $(I,J,G)$ families, sharing generated effects and errors, in the training data. Arithmetic mean reliability has the highest CV $R^2$ in eight of the nine combinations of inferential goal and estimator; $H$ has a small advantage for GR's ISEL. Figure~\ref{fig:metamodel} shows all nine comparisons, Table~\ref{tab:metamodel} gives three examples, and Table~\ref{tab:F-fair} gives the full numerical results. The small differences do not establish which summary would perform best for new families or trials. Proposition~2 concerns affine MSEL with known inputs and does not settle this comparison of fitted losses.

\FloatBarrier
\subsection{Estimator Performance Depends on the Goal}\label{subsec:sim-flips}

\begin{figure}[!htbp]
\centering
\includegraphics[width=\linewidth]{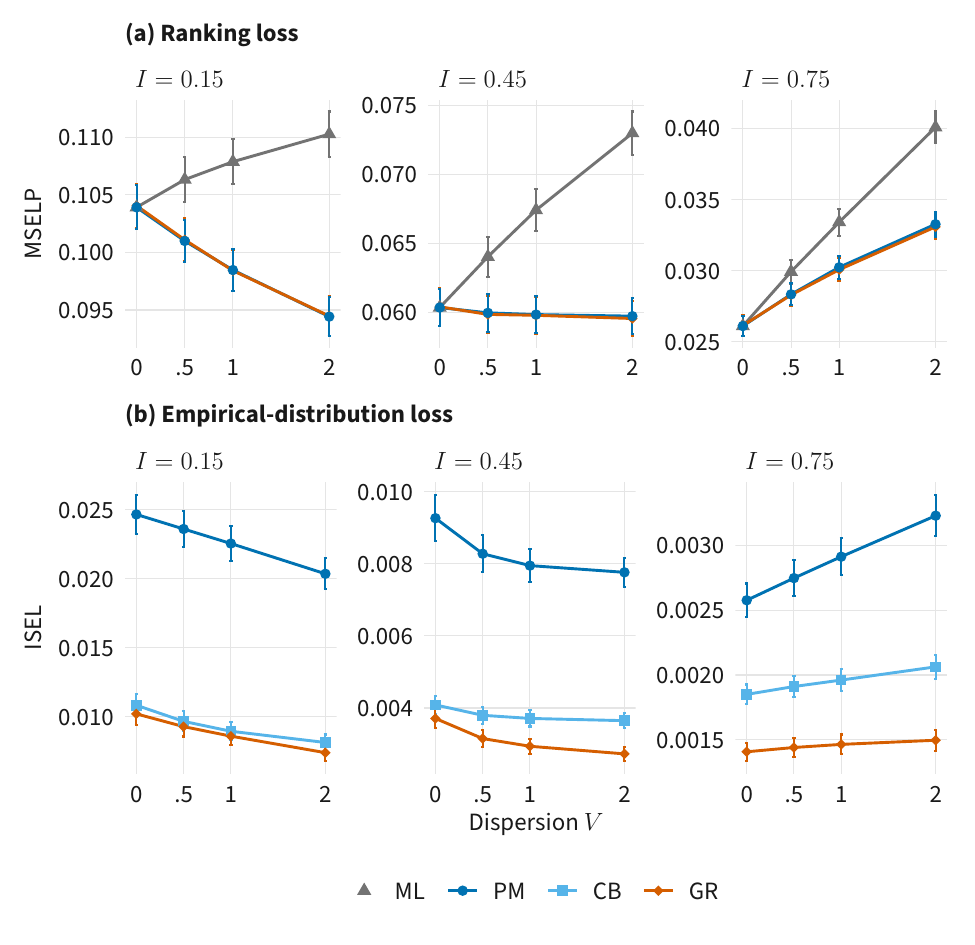}
\caption{Cell-mean ranking and EDF losses for $J=50$, Gaussian effects, and even profiles. Bars give the mean $\pm1.96$ MCSE for the implemented procedure. In (a), CB has the same rank loss as PM and is omitted; at $V=0$, both also share the raw estimates' ranks. Panel (b) shows the three shrinkage procedures. The intervals describe Monte Carlo uncertainty in mean losses, not site ranks.}
\label{fig:stress}
\end{figure}

For estimating individual site effects, PM has the lowest cell-mean RMSE among the four feasible estimators in all $702$ cells. Recovering the finite effect distribution leads to a different comparison: CB and GR have lower cell-mean ISEL than PM in $694$ and $688$ cells, respectively, with the largest gains for dispersed profiles at low $I$. Within GR, dispersion also changes distribution loss: at $I=.05$, its ratio for even profiles at $V=2$ relative to the homogeneous profile is $.572$ (MCSE $.007$).

For ranking, shrinkage makes no difference when every site receives the same increasing affine transformation. With homogeneous reliability, PM and the implemented CB estimator have this form and hence exactly the same ranks as the raw estimates. This equality also holds in the numerical results. With dispersion, GR can change the order of sites (Figure~\ref{fig:stress}). For even profiles at $V=2$, its geometric improvement in MSELP over the unpooled estimates is $8$--$17\%$ across levels of $I$.

The reliability profile therefore helps compare estimators for a specified goal. These comparisons concern average losses under the generating distributions studied here.

\FloatBarrier
\subsection{Imbalance under Fixed Total Enrollment}\label{subsec:sim-fixedN}

With total enrollment $N=J\bar n$ fixed, average $I$ decreases from $.333$ under balance to $.279$ at a parent Gamma CV of $.75$, while average $V$ increases to $.61$. PM's RMSE, ranking loss, and ISEL increase by about $3\%$, $8\%$, and $11\%$, respectively. These are arithmetic averages of cell-mean loss ratios relative to balance across the $27$ families, rather than the geometric paired ratios used in the fixed-$I$ comparisons.

For every estimator and loss, the aggregate ratios increase with the parent CV. Across the $27$ families, five procedures including the known-input linear benchmark, and three losses, $41$ of the $405$ sequences are nonmonotone; $14$ include a cell mean below balance.

Thus imbalance reduces informativeness and increases loss on average in this experiment. This agrees with Proposition~4, which makes balance optimal for $I$, $\bar S$, and affine MSEL with known inputs under the stated resource assumptions. The proposition does not cover every fitted estimator or loss. The apparent benefit of dispersion in some fixed-$I$ comparisons therefore does not imply a benefit from unequal allocation of a fixed total enrollment. Appendix~\ref{sec:F-simulation} gives further results by effect distribution, detailed ratios, and MCSEs for the fixed-$I$ summaries across the nine $J\times G$ families.

\section{Empirical Illustration: The Online Credit Recovery Study}\label{sec:ocrs}

\subsection{Data and Site-Level Estimates}\label{subsec:ocrs-data}

We use the Online Credit Recovery Study \citep{Rickles2024, AIR2024} to examine how much a completed trial can tell us about differences in effects across sites. We compare course passing with subject-matched credits earned by year four. In 2018 and 2019, first-year high school students who had failed Algebra 1 or English 9 were randomized to online summer credit recovery or the school's usual teacher-directed course. The study followed an earlier trial in Chicago \citep{Heppen2017}.

The public-use file contains $1{,}737$ student-by-subject records from $24$ schools. We define a site as a school-by-subject combination, giving $J=32$ sites with both treatment arms and $24$--$104$ records each (median $49$). Both outcomes are complete. For course passing, the effect is a risk difference.

Our site effects are block-adjusted regression contrasts. Within each site, we regress the outcome on treatment assignment and randomization-block fixed effects. Blocks containing both arms receive weights proportional to $n_kp_k(1-p_k)$, where $n_k$ is block size and $p_k$ is the treated proportion. Of the $144$ blocks, $22$ contain only one arm, representing $32$ records in $17$ sites. The regression weights differ from weights proportional to student counts in $28$ sites. These contrasts therefore generally differ from effects averaged over all students in a site. We also fit a covariate-adjusted model for course passing to assess sensitivity to this first-stage specification.

We compute HC2 sampling variances with a safeguard for the leverage denominator; its contribution to the treatment variance is negligible here (Appendix~\ref{subsec:G-estimation}). The hierarchy conditions on these estimated first-stage inputs. We obtain posterior means and intervals by numerical integration under the independent normal working model, with a flat reference prior on the mean and SD (Appendix~\ref{subsec:D-algorithm}).

\subsection{What the Diagnostics Show}\label{subsec:ocrs-dashboard}

\begin{figure}[!htbp]
\centering
\includegraphics[width=\linewidth]{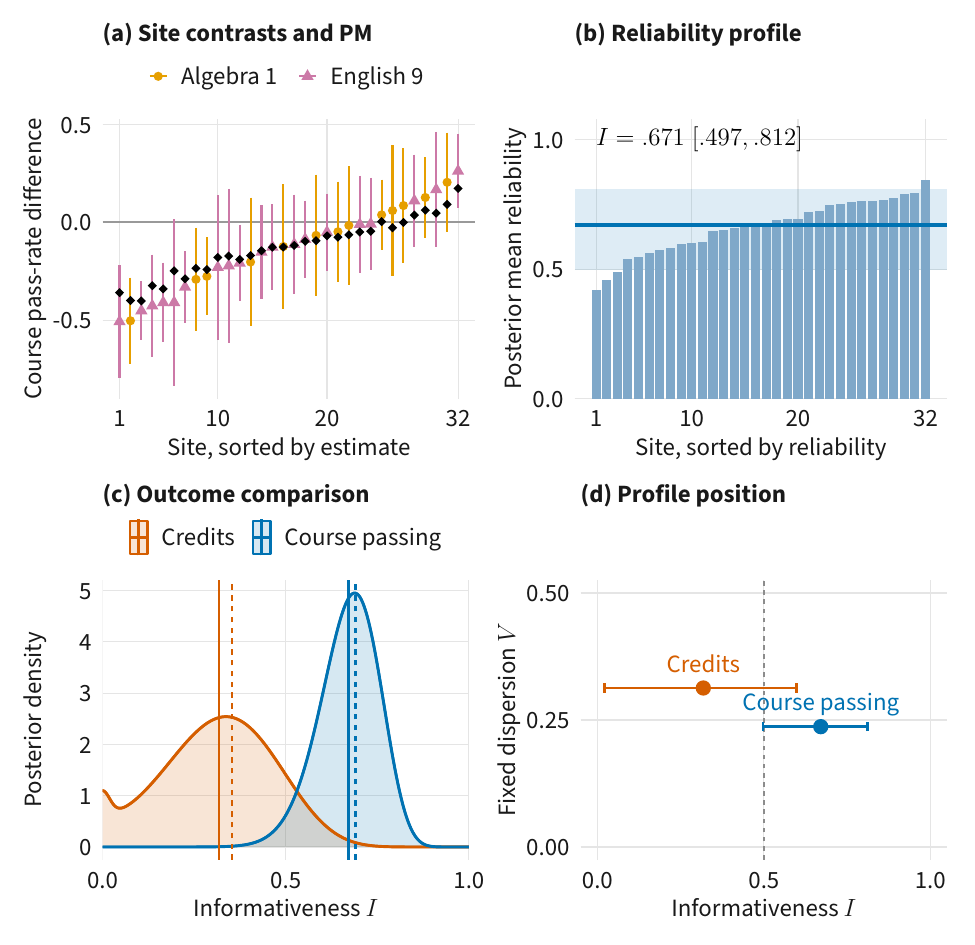}
\caption{Diagnostics for $32$ OCRS school-by-subject sites. (a) Course passing contrasts with first-stage normal 95\% intervals and posterior means (black diamonds). (b) Ordered posterior-mean reliabilities, with the posterior mean and 95\% band for $I$. (c) Smoothed posterior densities of $I$ for course passing and year-four credits. Solid lines indicate integrated means and dashed lines indicate plug-in summaries. (d) The $(I,V)$ profiles with 95\% posterior intervals for $I$. At $I=1/2$, signal variance equals geometric mean sampling variance; it also marks the sign change in the leading local term for $\bar S-I$ in Proposition~2. All summaries condition on the supplied variances. The integration method is described in Appendix~\ref{subsec:D-algorithm}.}
\label{fig:ocrs}
\end{figure}

The diagnostics indicate more information about site effects for course passing than for year-four credits (Table~\ref{tab:ocrs} and Figure~\ref{fig:ocrs}). For course passing, the posterior mean is $I=.67$, with a 95\% posterior interval of $[.50,.81]$ and $\Pr(I>1/2\mid\text{data})=.97$. For credits, informativeness is lower, $I=.32$ $[.02,.60]$, with $\Pr(I>1/2\mid\text{data})=.11$. These probabilities describe whether the signal variance exceeds the geometric mean sampling variance, since $I>1/2$ exactly when $\sigma^2>\tilde v$.

\begin{table}[!t]
\caption{Reliability profiles for the Online Credit Recovery Study.}
\label{tab:ocrs}
\centering
\begin{tablebody}
\begin{tabular}{@{}lccccc@{}}
\toprule
Analysis & $J$ & $I$ [95\% CrI] & $\bar S$ & $H$ [95\% CrI] & $J_{\mathrm{eff}}$ \\
\midrule
\multicolumn{6}{@{}l}{\emph{Course passing (risk difference)}} \\
\addlinespace[5pt]
Pooled & 32 & $.67\ [.50,.81]$ & $.66$ & $.025\ [.019,.029]$ & $31.2$ \\
\addlinespace[5pt]
Algebra 1 & 13 & $.63\ [.29,.87]$ & $.63$ & $.016\ [.009,.020]$ & $12.7$ \\
\addlinespace[5pt]
English 9 & 19 & $.71\ [.49,.87]$ & $.70$ & $.028\ [.018,.034]$ & $18.5$ \\
\addlinespace[5pt]
Adjusted & 32 & $.65\ [.47,.80]$ & $.64$ & $.028\ [.021,.032]$ & $31.0$ \\
\addlinespace[9pt]
\multicolumn{6}{@{}l}{\emph{Year-four credits (subject-matched)}} \\
\addlinespace[5pt]
Pooled & 32 & $.32\ [.02,.60]$ & $.33$ & $.028\ [.003,.038]$ & $28.8$ \\
\bottomrule
\end{tabular}
\end{tablebody}
\par\vspace{0.4\baselineskip}
\begin{minipage}{\linewidth}\normalsize\onehalfspacing\raggedright
\textit{Note.} Entries are posterior means of the functionals obtained by integration. Brackets give 95\% credible intervals conditional on the supplied variance estimates. The inputs are block-adjusted treatment contrasts; the adjusted analysis also includes the released test-score and GPA covariates and missingness indicators. $J_{\mathrm{eff}}$ is the effective site count defined in Section~\ref{subsec:jeff}. Fitted population summaries, the plug-in comparison, midpoint probability, and $V$ are given in Table~\ref{tab:G-ocrs-ancillary}. Appendix~\ref{sec:G-ocrs} describes the targets and sensitivity analyses.
\end{minipage}
\end{table}

The course passing profile also has little dispersion: $H=.025$, with arithmetic mean reliability $\bar S=.66$ and effective site count $J_{\mathrm{eff}}=31.2$ of $32$. The relatively high information level is therefore shared across sites rather than concentrated in a few. For credits, the effective count is $28.8$; its lower informativeness reflects generally lower reliabilities, with greater posterior uncertainty. These profiles describe the information available for site effects; accurate ranks and recovery of their finite distribution require separate assessment.

Evaluating informativeness at the posterior mean signal variance gives $.69$ for course passing, compared with $.67$ when we average informativeness over the posterior. The corresponding values for credits are $.35$ and $.32$. As Proposition~3 implies, the plug-in value is larger in every analysis. Table~\ref{tab:G-ocrs-ancillary} gives these comparisons, the fitted population mean, heterogeneity SD, and $V$. The subject-specific and covariate-adjusted results in Table~\ref{tab:ocrs} each refer to their own site collection or first-stage specification.

\subsection{Sensitivity to the Analysis Choices}\label{subsec:ocrs-sensitivity}

The contrast between outcomes persists when we retain only one site per school. Eight schools contribute both subjects, and $54$ students appear in both. Selecting one site from each of these schools gives $256$ collections, each with $J=24$. Across these collections, posterior-mean $I$ ranges from $.612$ to $.748$ for course passing and from $.116$ to $.421$ for credits. This checks sensitivity to the site collection; it does not correct joint uncertainty from dependence in the original $32$ sites.

Changing the first-stage contrast or effect scale changes informativeness. For course passing, $I$ is $.671$ with block-adjusted risk differences, $.632$ with unadjusted risk differences, and $.530$ with unadjusted log odds. The two unadjusted analyses use the same site-by-arm tables (Table~\ref{tab:G-ocrs-scale}). Risk differences and log odds depend differently on baseline risks, so their signal and sampling variances are not related by a common linear rescaling. The diagnostic therefore describes information about the specified effect measure.

We also refit the first stage and hierarchy in $500$ bootstrap samples drawn within fixed site-by-treatment-by-block cells. The refitted posterior means of $I$ average $.762$ for course passing and $.569$ for credits, with central 95\% ranges $[.665,.828]$ and $[.377,.726]$. Thus refitting changes location as well as spread, especially for credits, whose original posterior mean is $.318$. The cause of this shift remains unresolved. These results assess sensitivity to the resampling and refitting scheme. They do not validate the original conditional posterior intervals, establish coverage for the latent profile, or resolve shared-student dependence.

We impose variance floors in the refits to limit near-zero sampling variances. The credits floor never activates and cannot explain its shift. Halving or doubling the floors changes average refitted course passing $I$ by at most $.0011$. Dispersion is more sensitive: halving the floor moves the upper endpoint for $V$ from $.324$ to $.353$. Inferences about effects, ranks, or their distribution require their own uncertainty assessment \citep{LairdLouis1989, Lockwood2002}.

\section{Discussion}\label{sec:discussion}

We have shown that informativeness averages site reliabilities on the logit scale, explaining the geometric mean of sampling variances. It is also a Bregman centroid under Bernoulli--KL divergence. The minimum discrepancy from a constant profile under this loss gives the dispersion measure $H$. We chose this divergence to recover the existing index; the bounded range of reliability coefficients does not determine that choice.

When the common mean, effect variance, and sampling variances are known, arithmetic mean reliability $\bar S$ determines the minimum integrated MSEL among affine estimators based on a site's own estimate. We related this mean to informativeness through a derivative with centered logits held fixed and a local approximation that depends on profile shape. In the simulations, $\bar S$ generally predicted losses better than $H$, whose role is to quantify discrepancy under the loss defining $I$. Profiles with the same $I$ gave different estimation errors. PM had the lowest cell-mean RMSE among the feasible estimators, whereas CB and GR usually had lower loss for the finite effect distribution.

In the OCRS application, the working model indicated more information about site effects for course passing than for year-four credits. The course passing diagnostics also changed when we changed the first-stage contrast or effect measure. We recommend specifying the site-effect target, effect scale, first-stage estimator, and variance calculation when reporting these diagnostics. A common change of units multiplies signal and sampling variances by the same positive constant and leaves the diagnostics unchanged. Replacing risk differences with log odds ratios does not: both signal and noise change with baseline risks.

A trial with high informativeness can still contain imprecise sites: one sampling variance approaching zero can drive $I$ toward one while the others remain large. We therefore recommend plotting the reliability profile and reporting $I$ with $\bar S$ and either $H$ or logit variance $V$. $H$ measures discrepancy at the current information level, whereas $V$ describes the spread of log sampling variances and remains fixed as the common signal variance changes. The effective site count $J_{\mathrm{eff}}$ describes how evenly reliability is distributed across sites. These summaries provide no universal cutoffs or confidence guarantees for individual ranks \citep{Mogstad2024}. The R package \textsf{poolinfo} provides the diagnostics, posterior summaries, profile plots, and design calculations.

Varying dispersion at fixed $I$ generally changes both total precision and enrollment, so these comparisons do not answer how to allocate a fixed total enrollment. For a common inverse-size variance law and the fixed resources in Proposition~4, balanced allocation maximizes $I$ and $\bar S$. Choosing the number of sites, allowing unequal costs, or seeking accurate ranks requires a corresponding design criterion.

We computed posterior intervals under the normal working model, treating the estimated first-stage variances as known and hence holding $V$ fixed. These intervals should be reported with the model and prior. Refitting perturbed first-stage inputs assesses sensitivity to those inputs. The resulting variation differs from conditional posterior uncertainty. Both the normal and fixed-variance approximations may be poor with small or sparse first-stage samples. The geometric results do not supply a universal chi-square adjustment.

We also assumed independent sites. Selecting OCRS subsets changed the collection of sites and could not establish coverage for the original correlated collection. With known hyperparameters, jointly Gaussian sampling errors with covariance $\Omega$, and independent Gaussian prior effects, the full-vector shrinkage matrix is $\sigma^2(\sigma^2I_J+\Omega)^{-1}$. It generally differs from the diagonal rule based on scalar reliabilities. The marginal coefficient identities remain interpretable, but joint prediction and uncertainty require the covariance structure.

A model with a common mean and variance may be inadequate when site effects are related to sampling precision. Conditional location--scale modeling can address this association \citep{Chen2026}, although it changes the relation between model inputs and the scientific target. Our nonnormal simulations considered selected departures from the working model and do not establish general robustness for prediction, ranks, or EDF estimation. Further work is needed on the fitting procedures and inferential properties of flexible mixing distributions and models that allow covariance among site estimates.

\par\bigskip
\noindent\textbf{Data and software.}
The illustration uses the Online Credit Recovery Study public-use impact-analysis data from
\href{https://www.openicpsr.org/openicpsr/project/202181/version/V1/view}{openICPSR project 202181, version V1} \citep{AIR2024}, subject to the source's access and use terms.
Replication code, saved simulation results, and aggregate empirical inputs are available at
\url{https://github.com/joonho112/geometry-of-informativeness-replication}.
Participant-level data are not redistributed; the repository explains how to obtain them and repeat the first-stage analyses.
The accompanying R package \textsf{poolinfo} is available at
\url{https://github.com/joonho112/poolinfo}, with documentation and worked examples at
\url{https://joonho112.github.io/poolinfo/}.
Appendix~\ref{subsec:G-replication} describes the software and replication materials.

\smallskip
\noindent\textbf{Declaration of generative AI use.}
During the preparation of this work the author used Claude Opus 4.6 (Anthropic) for code-based checks and language editing. After using this tool, the author reviewed and edited the content as needed and takes full responsibility for the content of the published article.

\printbibliography[title={References}]

\clearpage
\appendix
\setcounter{section}{0}
\renewcommand{\thesection}{\Alph{section}}
\numberwithin{equation}{section}
\numberwithin{figure}{section}
\numberwithin{table}{section}
\renewcommand{\theHequation}{Supp.\thesection.\arabic{equation}}
\renewcommand{\theHfigure}{Supp.\thesection.\arabic{figure}}
\renewcommand{\theHtable}{Supp.\thesection.\arabic{table}}
\setcounter{theorem}{0}
\setcounter{proposition}{0}
\setcounter{corollary}{0}
\setcounter{lemma}{0}
\setcounter{definition}{0}

\phantomsection
\pdfbookmark[0]{Appendices}{appendices}
\begin{center}
{\Large\bfseries Appendices\par}
\vspace{1em}
{\large The Geometry of Informativeness: A Bregman-Centroid\\
Diagnostic for Partial Pooling in Multisite Trials\par}
\vspace{1.5em}
JoonHo Lee
\end{center}
\vspace{2em}
\startcontents[appendices]
\printcontents[appendices]{}{1}{\textbf{Contents}\vskip1em\hrule\vskip1em}
\vskip1em\hrule\vskip2em

\section{Projection Reliability and the Proof of Proposition~1}\label{sec:A-reliability}

\subsection{Conditional Expectation as Orthogonal Projection}\label{subsec:A-projection}

Let $X$ be square-integrable, and let $\mathcal G$ specify the information used to define its measurement target. In $L^2$, with inner product $\langle U,V\rangle=\E(UV)$, the conditional expectation $T=\E[X\mid\mathcal G]$ is the orthogonal projection onto the closed subspace of $\mathcal G$-measurable variables. The residual $E=X-T$ has mean zero and is orthogonal to this subspace \citep{Zumbo2026a}. Hence
\begin{equation}\label{eq:A-vardecomp}
\Var(X)=\Var(T)+\E(E^2).
\end{equation}
For $\Var(X)>0$, define $\mathrm{Rel}(X)=\Var(T)/\Var(X)$. If also $\Var(T)>0$, the centered variables $\widetilde X=X-\E X$ and $\widetilde T=T-\E T$ satisfy
\begin{equation}\label{eq:A-rel}
\mathrm{Rel}(X)=\frac{\Var(T)}{\Var(X)}
=\frac{\langle\widetilde X,\widetilde T\rangle^2}{\|\widetilde X\|^2\|\widetilde T\|^2}
=\cos^2\theta_{\widetilde X,\widetilde T}.
\end{equation}
The squared lengths of the centered score and its projection are their variances. Reliability is the fraction of the score's squared length retained by projection; a smaller angle means higher reliability. It differs from the norm of the projection operator, which equals one for every nonzero orthogonal projection.

\subsection{Proof of Proposition~1}\label{subsec:A-proof}

\begin{proof}
Set $X_j=\hat\tau_j$ and $\mathcal G_j=\sigma(\tau_j)$. By the conditional mean assumption, $T_j=\E[X_j\mid\tau_j]=\tau_j$. Let $\Var(\tau_j)=\sigma^2>0$ and $\E\{\Var(X_j\mid\tau_j)\}=v_j>0$. The law of total variance gives $\Var(X_j)=\sigma^2+v_j$, so
\[
\mathrm{Rel}(X_j)=\frac{\sigma^2}{\sigma^2+v_j}=S_j.
\]
Furthermore, $\langle\widetilde X_j,\widetilde T_j\rangle=\Var(\tau_j)$; substitution into~\eqref{eq:A-rel} gives $S_j=\cos^2\theta_j$ and $1-S_j=\sin^2\theta_j$. This sitewise identity needs neither normality nor independence between sites. Under the independent normal model with known hyperparameters, the same coefficient appears in the inverse posterior rule~\eqref{eq:posterior}. The posterior interpretation invokes the Gaussian model and does not follow from the projection identity alone.
\end{proof}

The conditional pooling coefficient $1-S_j$ on the hypermean differs in definition from the general posterior pooling measure of \citet{GelmanPardoe2006}. Proposition~1 does not equate these measures for arbitrary hierarchical fits.

\subsection{Boundary Cases and Assumptions}\label{subsec:A-boundary}

\begin{enumerate}[label=(\alph*)]
\item \emph{One vanishing sampling variance.} For fixed $\sigma^2>0$ and fixed positive $v_2,\ldots,v_J$, $v_1\downarrow0$ gives $S_1\to1$ and $I\to1$. For $J>1$, $V\to\infty$, whereas
\[
H\longrightarrow\frac1J\sum_{j\ne1}\ln(1/S_j).
\]
Thus the centroid can approach one even when some sites remain poorly measured. These limits are taken through positive inputs, since the logit construction excludes exact zero variances.
\item \emph{Vanishing between-site variance.} For fixed positive $v_j$, $\sigma^2\downarrow0$ gives $S_j,I,H\to0$. With a known common effect, there are then no departures between sites to estimate. The effective site count has the limit
\[
J_{\mathrm{eff}}\longrightarrow
\frac{(\sum_jv_j^{-1})^2}{\sum_jv_j^{-2}},
\]
although its defining ratio is $0/0$ at an exactly zero signal variance. The effective site count can therefore be large when all reliabilities are negligible; it describes their relative contributions.
\item \emph{Moment and input conditions.} Conditional unbiasedness, finite second moments, and the specified average conditional variance suffice for Proposition~1. The conditional variance can vary with $\tau_j$. The fitted model additionally assumes normality, independent sites, and a common effect distribution, and conditions on the supplied variance estimates. These assumptions require assessment when precision is informative about effects or the variance inputs are inaccurate. The moment identity alone does not ensure robustness to such departures.
\end{enumerate}

\subsection{Measurement Interpretation}\label{subsec:A-gtheory}

Generalizability theory emphasizes specifying the objects of measurement and relevant sources of error \citep{Cronbach1972,Brennan2001}. Here the object of measurement is the effect parameter specified for a site, and the observed score is its first-stage estimate. The signal and error variances must refer to the same target, whether defined by a joint probability law or by the sampling design. This interpretation provides a basis for a reliability summary. It does not imply that every multisite experiment has a particular crossed G-study design, determine the causal estimand, or establish the Gaussian inverse posterior rule.

\subsection{Conditional Complexity and Reliability Concentration}\label{subsec:A-complexity}

The effective site count describes how evenly reliability is distributed. Conditional complexity measures how strongly fitted site effects respond to the observations under specified conditioning assumptions.

Under the independent Gaussian model with $\tau,\sigma^2,v$ fixed, the derivative matrix of the posterior mean vector with respect to the observations is $\operatorname{diag}(S_1,\ldots,S_J)$, with trace $\sum_jS_j$. For the conditional likelihood of the site effects, defining $p_D$ in terms of posterior mean deviance also gives $\sum_j\Var(\tau_j\mid\hat\tau)/v_j=\sum_jS_j$ \citep{HodgesSargent2001,Spiegelhalter2002}.

If $\tau$ instead has a flat prior while $\sigma^2,v$ remain fixed, put $w_j=(\sigma^2+v_j)^{-1}$ and $W=\sum_jw_j$. The posterior mean of $\tau$ is $\sum_jw_j\hat\tau_j/W$. The site smoother consequently has trace
\[
\sum_jS_j+\frac{\sum_jw_j(1-S_j)}{W}
=\sum_jS_j+\frac{\sum_jS_j(1-S_j)}{\sum_jS_j}.
\]
The conditional $p_D$ has the same value, since the posterior variance for a site is $\sigma^2(1-S_j)+(1-S_j)^2/W$. With common $S_j=S$, the trace becomes $1+(J-1)S$. Estimating additional variance components, changing the likelihood, or allowing dependent observations requires a separate calculation. In general, these conditional parameter counts differ from the effective site count $J_{\mathrm{eff}}$, which depends only on the normalized reliability shares.

\section{Bregman Centroids and the Proof of Theorem~1}\label{sec:B-centroid}

\subsection{Bregman Divergences and Sided Centroids}\label{subsec:B-bregman}

Let $\phi$ be a real-valued, continuously differentiable, strictly convex potential on an open interval $\mathcal X\subseteq\mathbb R$. For $x,y\in\mathcal X$, define
\begin{equation}\label{eq:B-bregman}
D_\phi(x,y)=\phi(x)-\phi(y)-\phi'(y)(x-y).
\end{equation}
The divergence is the gap between the potential at the first argument and its tangent-line approximation from the second. Strict convexity gives $D_\phi(x,y)\geq0$, with equality exactly when $x=y$ \citep{Bregman1967,Banerjee2005}. The gap generally depends on argument order, so the two centroid problems need not have the same solution. Consider a finite profile $S_1,\ldots,S_J\in\mathcal X$ with uniform weights. Placing the candidate centroid in the second argument gives the \emph{right} centroid,
\begin{equation}\label{eq:B-right}
\operatorname*{arg\,min}_{c\in\mathcal X}\frac1J\sum_jD_\phi(S_j,c)
=\bar S=\frac1J\sum_jS_j.
\end{equation}
This follows by expanding the divergence, without differentiating $\phi'$ or assuming convexity in the second argument:
\[
\frac1J\sum_jD_\phi(S_j,c)
=\frac1J\sum_jD_\phi(S_j,\bar S)+D_\phi(\bar S,c).
\]
The final term has its unique minimum at $c=\bar S$. The conditional-expectation result for Bregman prediction uses this order under its assumptions on the random variables and their integrability \citep{BanerjeeGuoWang2005}.

Placing the candidate in the first argument gives the \emph{left} centroid,
\begin{equation}\label{eq:B-left}
I_\phi=\operatorname*{arg\,min}_{c\in\mathcal X}\frac1J\sum_jD_\phi(c,S_j).
\end{equation}
Here $\phi'$ is continuous and strictly increasing. Convexity makes the derivative nondecreasing. If it had the same value at two distinct points, $\phi$ would be affine between them, contradicting strict convexity. The finite average of the $\phi'(S_j)$ lies between their minimum and maximum and is therefore in the image of $\phi'$. The objective in~\eqref{eq:B-left} equals $\phi(c)-cJ^{-1}\sum_j\phi'(S_j)$ plus a constant. Its derivative changes sign at the unique point satisfying
\begin{equation}\label{eq:B-dual}
\phi'(I_\phi)=\frac1J\sum_j\phi'(S_j),
\qquad
I_\phi=(\phi')^{-1}\!\left\{\frac1J\sum_j\phi'(S_j)\right\}.
\end{equation}
To obtain this centroid, we transform the profile values, take their arithmetic mean on that scale, and transform back. A unique interior minimizer exists without requiring twice differentiability or an everywhere positive second derivative. We use these classical formulas for sided centroids \citep{NielsenNock2009,NielsenNock2020} to identify the profile objective minimized by an existing multisite index.

\subsection{The Bernoulli--KL Geometry on \texorpdfstring{$(0,1)$}{(0,1)}}\label{subsec:B-bernoulli}

For $\mathcal X=(0,1)$, take the convex negative Bernoulli entropy
\begin{equation}\label{eq:B-potential}
\begin{gathered}
\phi_{\mathrm{KL}}(p)=p\ln p+(1-p)\ln(1-p),\\
\phi_{\mathrm{KL}}'(p)=\ln\frac{p}{1-p}=\operatorname{logit}(p),
\qquad
\phi_{\mathrm{KL}}''(p)=\frac1{p(1-p)}.
\end{gathered}
\end{equation}
Then
\begin{equation}\label{eq:B-klform}
D_{\phi_{\mathrm{KL}}}(p,q)
=p\ln\frac pq+(1-p)\ln\frac{1-p}{1-q}
=\KL\{\Ber(p)\|\Ber(q)\}.
\end{equation}
For this Bernoulli family, the Fisher information in the mean parameter $p$ is $1/\{p(1-p)\}=\phi_{\mathrm{KL}}''(p)$. In the natural parameter $\eta=\operatorname{logit}(p)$, the log-partition function is $A(\eta)=\ln(1+e^\eta)$ and the Fisher information is $A''(\eta)=p(1-p)$ \citep{Amari2016}. The two expressions use different coordinates. The Bernoulli embedding defines a loss for comparing reliability coefficients; it makes no assumption about the sampling distribution of the site-effect estimates. Our objective concerns a finite profile; psychometric Bregman formulations project random scores \citep{Zumbo2026b}.

\subsection{Proof of Theorem~1}\label{subsec:B-proof}

With a common signal variance, averaging logit reliabilities subtracts the average log sampling variance from a constant. Transforming back therefore introduces a geometric mean.

\begin{proof}
With common $\sigma^2>0$ and finite $v_j>0$, every $S_j=\sigma^2/(\sigma^2+v_j)$ lies in $(0,1)$. Applying~\eqref{eq:B-dual} to~\eqref{eq:B-potential} gives the unique interior left centroid, satisfying
\begin{equation}\label{eq:B-foc}
\operatorname{logit}(I)=\frac1J\sum_j\operatorname{logit}(S_j).
\end{equation}
Because $\operatorname{logit}(S_j)=\ln\sigma^2-\ln v_j$, inversion yields
\[
I=\frac{\sigma^2}{\sigma^2+\exp\{J^{-1}\sum_j\ln v_j\}}
=\frac{\sigma^2}{\sigma^2+\tilde v},
\]
which is the index in~\eqref{eq:informativeness}. This identification holds for fixed variance inputs and uniform weights and does not require normal sampling.
\end{proof}

\subsection{A Conditional Characterization of the Generator}\label{subsec:B-characterization}

We characterize the fixed potentials that reproduce the logit mean for every interior profile. Write $\ell(t)=(1+e^{-t})^{-1}$.

\begin{lemma}[Logit-mean matching for every profile]\label{lem:B-logit-characterization}
Let $\phi$ be a fixed real-valued $C^1$, strictly convex scalar function on $(0,1)$, used for every profile. Define the two-point centroid with the candidate in the first argument by
\[
c_\phi(p,q)=\operatorname*{arg\,min}_{c\in(0,1)}
\tfrac12\{D_\phi(c,p)+D_\phi(c,q)\}.
\]
Then
\[
c_\phi(p,q)=\ell\!\left(\frac{\operatorname{logit}(p)+\operatorname{logit}(q)}2\right)
\]
for every $p,q\in(0,1)$ if and only if
\[
\phi(p)=\alpha\{p\ln p+(1-p)\ln(1-p)\}+\beta p+\gamma,
\qquad \alpha>0,
\]
for constants $\alpha,\beta,\gamma$. In that case $D_\phi=\alpha D_{\phi_{\mathrm{KL}}}$ and the matching holds for every finite uniformly weighted interior profile.
\end{lemma}

\begin{proof}
Let $g(t)=\phi'(\ell(t))$, a continuous strictly increasing function on $\mathbb R$. By~\eqref{eq:B-dual}, matching the two-point centroids gives the midpoint equation
\[
g((u+v)/2)=\{g(u)+g(v)\}/2,\qquad u,v\in\mathbb R.
\]
Set $h(t)=g(t)-g(0)$. Taking the pair $(2t,0)$ gives $h(2t)=2h(t)$; combining this with the midpoint equation gives $h(u+v)=h(u)+h(v)$. Rational homogeneity and continuity then yield $h(t)=\alpha t$ for all real $t$, with $\alpha=h(1)>0$ by strict increase. Thus $\phi'(p)=\alpha\operatorname{logit}(p)+\beta$, where $\beta=g(0)$. Integration gives the stated form with an arbitrary constant $\gamma$. Conversely, this derivative in~\eqref{eq:B-dual} gives the logit mean for every finite uniform profile. Affine terms cancel from the divergence, leaving $D_\phi=\alpha D_{\phi_{\mathrm{KL}}}$.
\end{proof}

For any fixed $\sigma^2>0$, each interior $p$ can be written as $\sigma^2/(\sigma^2+v)$ by taking $v=\sigma^2(1-p)/p>0$. The premise therefore requires the centroids to agree for unrestricted positive variance profiles; agreement in one observed trial or a finite design grid is insufficient. Positive rescaling preserves the centroid while changing the units of divergence. We specify these units by using natural logarithms and $\alpha=1$. The characterization concerns this centroid rule and does not establish which estimation loss is empirically optimal.

\subsection{Remarks}\label{subsec:B-remarks}

\begin{enumerate}[label=(\alph*)]
\item \emph{Order of arguments.} The left centroid minimizes the average $D_\phi(c,S_j)$, whereas the right centroid minimizes the average $D_\phi(S_j,c)$. For the Bernoulli potential these centroids are $I$ and $\bar S$, respectively. A centroid for a finite profile and a conditional predictor solve different optimization problems even when they have the same formula.
\item \emph{Scale invariance.} Multiplying all of $(\sigma^2,v_1,\ldots,v_J)$ by a common positive constant leaves the reliabilities, $I$, $H$, $V$, and $J_{\mathrm{eff}}$ unchanged. Rescaling the divergence generator instead changes $H$.
\item \emph{Centroid ordering.} Set $r_j=(1-S_j)/S_j>0$. The harmonic centroid satisfies
\[
\operatorname{HM}(S)=\frac1{1+\operatorname{AM}(r)}
\leq\frac1{1+\operatorname{GM}(r)}=I,
\]
and $\operatorname{HM}(S)\leq\bar S$ by the harmonic--arithmetic mean inequality. Equality in either inequality requires a constant profile. There is no general ordering of $I$ and $\bar S$, as the skewed-profile example in Appendix~\ref{sec:C-dispersion} shows. For $(S_1,S_2)=(.2,.8)$, both centroids are $.5$, but $H=\ln(5/4)>0$. Equality of the centroids therefore does not imply homogeneity.
\item \emph{Endpoint limits.} If one $S_j$ approaches $0$ or $1$ while all others remain fixed and interior, the mean logit diverges and $I$ approaches that boundary; Appendix~\ref{sec:A-reliability} discusses the interpretation for the profile. Approaching opposite endpoints can produce a different limit: $S_1=\epsilon$, $S_2=1-\epsilon$ gives $I=1/2$ for every $0<\epsilon<1$. If one coefficient is exactly $0$ and another exactly $1$, the average KL objective with the candidate in the first argument is infinite for every interior candidate. The joint limit can depend on the path, so the interior formulas do not define a unique centroid for this exact endpoint profile.
\end{enumerate}

\section{Dispersion: Proofs of Theorem~2 and Proposition~2}\label{sec:C-dispersion}

\subsection{Proof of Theorem~2}\label{subsec:C-pythagoras}

\begin{proof}
Let $\mathcal X$ be an open interval, let $\phi\in C^1(\mathcal X)$ be strictly convex, and let $S_j\in\mathcal X$. The left centroid $I_\phi$ exists and satisfies~\eqref{eq:B-dual}, as shown in Appendix~\ref{sec:B-centroid}. For every $p\in\mathcal X$, expansion gives
\begin{align}
\frac1J\sum_jD_\phi(p,S_j)
&=\phi(p)-\frac1J\sum_j\phi(S_j)
-\frac1J\sum_j\phi'(S_j)(p-S_j),\label{eq:C-lhs}\\
D_\phi(p,I_\phi)+\frac1J\sum_jD_\phi(I_\phi,S_j)
&=\phi(p)-\phi'(I_\phi)(p-I_\phi)-\frac1J\sum_j\phi(S_j)\nonumber\\
&\quad-\frac1J\sum_j\phi'(S_j)(I_\phi-S_j).\label{eq:C-rhs}
\end{align}
The difference is
\begin{equation}\label{eq:C-gap}
\frac1J\sum_jD_\phi(p,S_j)-\{D_\phi(p,I_\phi)+H_\phi\}
=(p-I_\phi)\left\{\phi'(I_\phi)-\frac1J\sum_j\phi'(S_j)\right\}=0.
\end{equation}
Since $D_\phi(p,I_\phi)$ is nonnegative and vanishes only at the centroid, $H_\phi$ is the minimum of this objective and is attained only at $p=I_\phi$.

For the converse, suppose the identity holds for all $p$ with center $c\in\mathcal X$ and constant $h$. Setting $p=c$ gives $h=J^{-1}\sum_jD_\phi(c,S_j)$. Applying the same expansion at $c$ leaves $(p-c)\{\phi'(c)-J^{-1}\sum_j\phi'(S_j)\}=0$ for every $p$. Taking $p\ne c$ and using strict monotonicity of $\phi'$ gives $c=I_\phi$, and hence $h=H_\phi$.
\end{proof}

As in the sum-of-squares decomposition about a mean, the cross term vanishes at the centroid. Here we compare a common candidate value across sites under the specified divergence. This identity does not require a projection equality for arbitrary convex sets, and the residual divergence $H_\phi$ need not be a squared metric distance. The potential, argument order, and weights determine the discrepancy.

For Bernoulli negative entropy, write $\eta_j=\phi_{\mathrm{KL}}'(S_j)=\operatorname{logit}(S_j)$ and $m=J^{-1}\sum_j\eta_j$. Its conjugate is $\phi_{\mathrm{KL}}^*(t)=A(t)=\ln(1+e^t)$, and Legendre duality reverses the divergence arguments \citep{Amari2016}:
\[
D_{\phi_{\mathrm{KL}}}(I,S_j)=D_A(\eta_j,m).
\]
Consequently,
\[
H=\frac1J\sum_jA(\eta_j)-A(m),
\]
which is the Jensen gap in dual coordinates: the average of the log-partition function across the site logits minus its value at their mean. The minimum for the right centroid is $J^{-1}\sum_j\phi_{\mathrm{KL}}(S_j)-\phi_{\mathrm{KL}}(\bar S)$ and can differ from the minimum for the left centroid \citep{NielsenNock2020}. For example, $(S_1,S_2)=(.2,.8)$ gives $I=\bar S=.5$, but the left minimum is $\ln(5/4)\approx.223144$, whereas the right minimum is approximately $.192745$.

\subsection{Proof of Proposition~2}\label{subsec:C-local}

Write $s(t)=(1+e^{-t})^{-1}$, so that $I=s(m)$ and $S_j=s(m+\Delta_j)$. Let $\overline\Delta=0$, $V=J^{-1}\sum_j\Delta_j^2$, and $\rho_3=J^{-1}\sum_j|\Delta_j|^3$.

\begin{proof}
\emph{Part (i).} Direct expansion of each Bernoulli divergence gives
\[
\mathrm{KL}\{\Bern(I)\|\Bern(S_j)\}
=-I\Delta_j+A(m+\Delta_j)-A(m).
\]
Averaging cancels the linear term, so
\begin{equation}\label{eq:C-Hexact}
H(m;\Delta)=\frac1J\sum_jA(m+\Delta_j)-A(m).
\end{equation}
Since $A'=s$, differentiation with $\Delta$ held fixed yields $\partial H/\partial m=\bar S-I$. Using $dm/dI=\{I(1-I)\}^{-1}$, the derivative with respect to $I$ is $(\bar S-I)/\{I(1-I)\}$ along the same path.

\emph{Part (ii).} Expand $A(m+\Delta_j)$ to second order about $m$. The constant and linear terms cancel in~\eqref{eq:C-Hexact}, and $A''(m)=I(1-I)$. For $u=s(t)$,
\[
A'''(t)=u(1-u)(1-2u),\qquad
\sup_t|A'''(t)|=\frac{\sqrt3}{18}.
\]
Taylor's theorem with Lagrange remainder therefore gives
\begin{equation}\label{eq:C-Hexp}
H=\tfrac12I(1-I)V+R_H,\qquad
|R_H|\le\frac{\sqrt3}{108}\rho_3.
\end{equation}
The curvature $A''(m)=I(1-I)$ is the Bernoulli Fisher information in the natural (logit) coordinate. In the probability coordinate it is $\phi_{\mathrm{KL}}''(I)=1/\{I(1-I)\}$.

Next expand $s(m+\Delta_j)$ to second order. Because $s''(m)=I(1-I)(1-2I)$, $\overline\Delta=0$, and
\[
s'''(t)=u(1-u)(1-6u+6u^2),\qquad
\sup_t|s'''(t)|=\frac18,
\]
we obtain
\begin{equation}\label{eq:C-Sbar}
\bar S-I=\tfrac12I(1-I)(1-2I)V+R_S,\qquad
|R_S|\le\frac{\rho_3}{48}.
\end{equation}
Combining the expansions also gives
\begin{equation}\label{eq:C-gap-remainder}
\bar S-I=(1-2I)H+R,\qquad
|R|\le\left(\frac1{48}+|1-2I|\frac{\sqrt3}{108}\right)\rho_3,
\end{equation}
where $R=R_S-(1-2I)R_H$. The bound holds for every finite interior profile, but the leading term is accurate only if the remainder is sufficiently small.

\emph{Part (iii).} Set $a=\sigma^2$, $Y_j=\hat\tau_j$, and $\mu=\tau$. With known common mean $E(\tau_j)=\mu$, the conditional-moment assumptions of Proposition~1 give
\[
E(Y_j)=E(\tau_j)=\mu,\quad
\operatorname{Var}(Y_j)=a+v_j,\quad
\operatorname{Cov}(Y_j,\tau_j)=a.
\]
For a predictor affine in $Y_j$, optimizing the intercept gives $\delta_j=\mu+b_j(Y_j-\mu)$. Its integrated squared-error risk is
\begin{align*}
E\{(\delta_j-\tau_j)^2\}
&=a-2b_ja+b_j^2(a+v_j)\\
&=\frac{av_j}{a+v_j}+(a+v_j)(b_j-S_j)^2.
\end{align*}
Thus $b_j=S_j$ minimizes risk among affine predictors using only that site's estimate. The minimum is $a(1-S_j)$; averaging with equal site weights gives
\[
\mathcal R_{\mathrm{aff}}=a(1-\bar S)
=a\left\{1-I-\frac{\partial H}{\partial m}\right\}.
\]
The sum of marginal risks does not require independence between sites. Under the independent Gaussian model with known $\mu,a,v_j$, $\delta_j$ is also the posterior mean given all observations, and $\operatorname{Var}(\tau_j\mid Y)=a(1-S_j)$ \citep{Morris1983}. In that model, the expression therefore gives both posterior expected MSEL and its prior-predictive average.

Substituting~\eqref{eq:C-gap-remainder} into the expression for risk gives
\begin{align*}
\mathcal R_{\mathrm{aff}}
&=a\{(1-I)-(1-2I)H\}+R_{\mathrm{risk}},\\
|R_{\mathrm{risk}}|
&\le a\left(\frac1{48}+|1-2I|\frac{\sqrt3}{108}\right)\rho_3,
\end{align*}
where $R_{\mathrm{risk}}=-aR$. The effect variance $a$ enters the remainder as a multiplicative factor. The $O(\rho_3)$ bound on the risk error therefore depends on this scale.
\end{proof}

\subsection{Discussion of the Expansion}\label{subsec:C-remarks}

\emph{Fixed inputs and differentiation.} Since $\Delta_j=\overline{\ln v}-\ln v_j$, both $\Delta$ and $V$ remain fixed when $a$ varies with the $v_j$ held fixed. All site logits then move by the same amount: the location changes, but the centered shape does not. Uncertainty in estimated first-stage variances changes the profile and requires separate treatment. The derivative applies to this path; it does not describe arbitrary perturbations of the variances, allocation under a fixed budget, or the total risk of a fitted model.

\emph{Local sign and a skewed counterexample.} The leading term in~\eqref{eq:C-Sbar} has the sign of $1-2I$. A sufficient condition for that sign to hold is
\[
\left|\tfrac12I(1-I)(1-2I)V\right|>\rho_3/48.
\]
Consider a fixed nonconstant centered shape and a fixed $I\ne1/2$. As the shape is scaled towards zero, the second-order term eventually dominates the third-order bound. This local result does not establish the midpoint rule for arbitrary profiles or ensure its validity over ranges encountered in practice.

Take $J=25$, $m=\operatorname{logit}(.30)$, and
\[
\Delta_1=\sqrt{48},\qquad
\Delta_2=\cdots=\Delta_{25}=-\sqrt{1/12}.
\]
Then $\overline\Delta=0$ and $V=2$ exactly, but $\bar S=.2732468357<I$. Replacing $m$ by $-m$ and each $\Delta_j$ by $-\Delta_j$ gives $I=.70$ and $\bar S=.7267531643>I$. In both cases, comparison with a homogeneous profile at the same $I$ and $a$ reverses the conclusion suggested by the leading midpoint rule. Nor does equality of the centroids imply homogeneity: $(.2,.8)$ has $I=\bar S=.5$ and positive $H$. Consequently, the gap, $H$, and $V$ cannot be used interchangeably as monotone measures.

\emph{Interpretation of risk.} The affine calculation assumes the specified joint moments and known $\mu,a,v_j$. Under the normal sampling model, conditioning on a fixed realized effect gives
\[
E\{(\delta_j-\tau_j)^2\mid\tau_j\}
=S_j^2v_j+(1-S_j)^2(\tau_j-\mu)^2,
\]
which generally differs from $a(1-S_j)$. The integrated risk is recovered only by averaging over the specified effect distribution. Outside the independent Gaussian model, the optimal affine predictor need not be optimal among nonlinear predictors or predictors using the full vector of observations. Estimating hyperparameters requires a separate risk calculation. The simulations compare fitted procedures under the stated losses. The reference $a(1-I)$ is the risk for a homogeneous profile. It does not give the risk of applying a common slope $I$ to the actual heterogeneous profile.

\section{Estimation of the Diagnostics: Proof of Proposition~3 and Computation}\label{sec:D-estimation}

\subsection{Proof of Proposition~3}\label{subsec:D-proof}

With sampling variances fixed, informativeness increases with heterogeneity variance at a decreasing rate. This curvature explains the averaging comparison in Proposition~3.

\begin{proof}
Fix $c=\tilde v>0$ and let $a=\sigma^2\ge0$. The derivatives
\[
I'(a)=\frac{c}{(a+c)^2}>0,
\qquad I''(a)=-\frac{2c}{(a+c)^3}<0
\]
establish strict concavity on $[0,\infty)$. Jensen's inequality applies when the posterior is proper and has finite mean $m=\E(a\mid\text{data})$. By strict concavity, equality holds exactly when the posterior is a point mass. Replacing $c$ by $v_j$ proves the result for each site.
\end{proof}

Because the index is bounded, its posterior mean $\E\{I(a)\mid\text{data}\}$ exists for any proper posterior, even if $m$ is infinite. The plug-in comparison requires a finite mean. Suppressing the conditioning in the notation, the identity
\[
\frac1{a+c}=\frac1{m+c}-\frac{a-m}{(m+c)^2}
+\frac{(a-m)^2}{(m+c)^2(a+c)}
\]
gives an exact Jensen gap whenever $m<\infty$:
\begin{equation}\label{eq:D-jensen-gap}
G_I:=I(m)-\E\{I(a)\}
=\frac{c}{(m+c)^2}\E\left\{\frac{(a-m)^2}{a+c}\right\}\ge0.
\end{equation}
The expectation is finite under the assumed finite mean because its integrand grows only linearly as $a\to\infty$.

If the posterior variance $s_a^2$ is finite, a further exact expansion yields
\[
G_I=\frac{c\,s_a^2}{(m+c)^3}+R,
\qquad
R=-\frac{c}{(m+c)^3}\E\left\{\frac{(a-m)^3}{a+c}\right\}.
\]
The expectation defining $R$ requires only the second moment. If the third absolute central moment $\rho_a=\E|a-m|^3$ is also finite, then
\[
|R|\le\frac{\rho_a}{(m+c)^3}.
\]
The approximation $G_I\approx\tfrac12|I''(m)|s_a^2$ requires finite variance and a remainder small relative to the leading term; the bound can verify the latter condition. Finite moments alone ensure neither accuracy nor a rate in $J$. These posterior comparisons at fixed data do not establish frequentist bias or coverage properties.

\subsection{Reference Prior and Posterior Moments}\label{subsec:D-prior}

We use an improper reference prior that is flat with respect to $d\tau\,d\sigma$ on $\mathbb R\times[0,\infty)$. When expressed in terms of $a=\sigma^2$, its density with respect to $d\tau\,da$ is proportional to $a^{-1/2}$ for $a>0$ and hence is not flat in the variance. A finite cutoff in the numerical calculation approximates this reference analysis rather than specifying a fixed proper uniform prior. Proper priors for the heterogeneity standard deviation require a choice of scale and a corresponding sensitivity assessment \citep{Gelman2006}.

A proper posterior need not have the finite moments required for the preceding comparisons. Let $y_j=\hat\tau_j$ be finite and $0<v_j<\infty$ be fixed, and assume the independent marginal working likelihood $y_j\mid\tau,\sigma\sim\mathsf N(\tau,\sigma^2+v_j)$. Write
\[
w_j(\sigma)=(\sigma^2+v_j)^{-1},\qquad
W(\sigma)=\sum_jw_j(\sigma),\qquad
\hat\mu_\sigma=\frac{\sum_jw_j(\sigma)y_j}{W(\sigma)}.
\]
Completing the square gives
\[
\sum_jw_j(y_j-\tau)^2
=\sum_jw_j(y_j-\hat\mu_\sigma)^2+W(\tau-\hat\mu_\sigma)^2.
\]
Integrating the final Gaussian factor over the flat mean prior yields the unnormalized marginal posterior density
\begin{equation}\label{eq:D-marginal-sigma}
g(\sigma)=W(\sigma)^{-1/2}
\prod_j(\sigma^2+v_j)^{-1/2}
\exp\left\{-\frac12\sum_jw_j(\sigma)(y_j-\hat\mu_\sigma)^2\right\}.
\end{equation}
As $\sigma\downarrow0$, the fixed positive $v_j$ ensure that $g$ remains finite and positive. As the standard deviation tends to infinity, $W(\sigma)\sim J/\sigma^2$, the product is asymptotic to $\sigma^{-J}$, and the exponential term tends to one. Hence
\[
g(\sigma)\sim J^{-1/2}\sigma^{1-J}.
\]
The integral at infinity is finite exactly when $J>2$. If the posterior is proper, then for any $r\ge0$,
\[
\E(a^r\mid\text{data})<\infty
\quad\Longleftrightarrow\quad J>2r+2,
\]
because the moment integrand has tail proportional to $\sigma^{2r+1-J}$. Equality at the threshold gives a logarithmically divergent integral. In particular, the mean of $a$, its variance, and its third moment require $J>4$, $J>6$, and $J>8$, respectively.

The numbers of sites in the simulations, $J=25,50,100$, and in the application or its site subsets, $J=32,24,13,19$, satisfy these moment conditions under the specified likelihood and prior with positive variance inputs. These conditions establish the existence of moments; numerical convergence and the assumptions of independence and normality must be assessed separately. Separate derivations are also required for known means, other priors, zero supplied variances, or correlated sampling errors.

\subsection{Computation of Posterior Functionals}\label{subsec:D-algorithm}

We average each diagnostic over uncertainty in heterogeneity, holding the supplied sampling variances fixed. For a diagnostic $f(\sigma;v)$, the target posterior mean is
\[
\E\{f(\sigma;v)\mid\text{data}\}
=\frac{\int_0^\infty f(\sigma;v)g(\sigma)\,d\sigma}
{\int_0^\infty g(\sigma)\,d\sigma},
\]
with $g$ given by~\eqref{eq:D-marginal-sigma}. We obtain $f$ from $S_j=\sigma^2/(\sigma^2+v_j)$ and the definitions of $I,H,\bar S$, and $J_{\mathrm{eff}}$. The variance of the logits, $V$, is constant conditional on the supplied variance vector but changes when that vector is re-estimated.

\paragraph{Application and sensitivity summaries.}
For the application, we compute posterior means by adaptive quadrature over $[0,\infty)$, splitting the integral at $\sqrt{\tilde v}$ and rescaling the density. The relative integration tolerance is $10^{-10}$, and the absolute tolerance for the rescaled integrals is $10^{-12}$. We also integrate on the log-$\sigma$ scale, including the Jacobian, and obtain normalizing constants and diagnostic means that agree to numerical precision. The same procedure gives the posterior means for the bootstrap refits and site subsets in Appendix~\ref{sec:G-ocrs}. Percentiles across these means summarize variation among the refits.

For finite numerical representations, we bound the integrals over the entire omitted tail. For $\sigma\ge B>0$, $W(\sigma)\ge J/(\sigma^2+v_{\max})$ implies
\[
g(\sigma)\le\frac{\sqrt{1+v_{\max}/B^2}}{\sqrt J}\sigma^{1-J}.
\]
Integrating this envelope against $\sigma^q$, for $q<J-2$, bounds the unnormalized tail moment by
\[
\frac{\sqrt{1+v_{\max}/B^2}}{\sqrt J}
\frac{B^{q+2-J}}{J-q-2}.
\]
We evaluate these bounds for $q=0,2,4$ using the numerical normalizing constant. The finite representations used to cache draws have normalized mass bounds below $10^{-12}$ and moment bounds below $10^{-10}\tilde v^{q/2}$ for $q=2,4$. The adaptive integrals themselves extend to infinity. The tail bounds, integration-error estimates, and comparisons between calculations assess numerical approximation; calibration of the working posterior remains a separate question.

Posterior intervals for $I$, $\bar S$, and $J_{\mathrm{eff}}$ follow from monotone transformations of the integrated $\sigma$ CDF. For $J_{\mathrm{eff}}$, fix positive $v_j$, let $a=\sigma^2>0$, and write $A_k(a)=\sum_j(a+v_j)^{-k}$. Since $S_j=a/(a+v_j)$,
\[
J_{\mathrm{eff}}(a)=\frac{A_1(a)^2}{A_2(a)},\qquad
\frac{d}{da}\log J_{\mathrm{eff}}(a)
=2\left\{\frac{A_3(a)}{A_2(a)}-\frac{A_2(a)}{A_1(a)}\right\}\ge0,
\]
where the inequality follows from $A_2(a)^2\le A_1(a)A_3(a)$ by Cauchy--Schwarz. The increase is strict unless all $v_j$ are equal, in which case $J_{\mathrm{eff}}=J$ and its posterior interval is $[J,J]$.

To obtain intervals for $H$, we partition the $\sigma$ domain at the zeros of $\bar S-I$, invert the monotone branches, and sum their posterior probabilities. We therefore do not obtain its two tails simply by transforming the two endpoint quantiles of $\sigma$. Searches for turning points at two resolutions agreed for the profiles examined; homogeneous profiles have $H=0$ identically. We independently integrated the direct Bernoulli--KL formula for the course passing and credits means and the $I/H$ quantiles, obtaining agreement within $3.5\times10^{-13}$.

For effect draws and density displays, we retain a cache of $20{,}000$ draws. We draw $\sigma$ using an interpolated inverse CDF, followed by
\[
\tau\mid\sigma,\text{data}\sim\mathsf N\{\hat\mu_\sigma,W(\sigma)^{-1}\}
\]
and the conditional site-effect draws in~\eqref{eq:posterior}. The cached inverse CDF agrees with the integrated CDF within $1.1\times10^{-8}$ at the probability points examined. To obtain posterior means and variances for the site effects, we integrate the conditional hypermean analytically and then integrate over $\sigma$. We use quadrature for the reported diagnostic means and intervals. For stable density displays, we evaluate the inverse CDF at midpoint probabilities to obtain deterministic samples, distinct from the random posterior draws.

\paragraph{Simulation computation.}
The main and fixed-enrollment simulations use $250$ grid points and $M=2{,}000$ draws per fit. The grid starts with upper bound
\[
b=\max\left\{3\sqrt{\max(0,s_y^2-\bar v)},\ 2s_y,\ 3\sqrt{\operatorname{median}(v_j)},\ 10^{-3}\right\},
\]
where $s_y$ is the sample standard deviation of the $y_j$. We shift the zero node to $b/(10\times250)$ and normalize the density values at the nodes. We double the upper bound until the normalized mass on nodes above $0.95b$ is below $10^{-4}$ or expansion stops at the safeguard $b>100s_y$. Sampled nodes receive uniform jitter on $[-\delta/2,\delta/2]$, where $\delta$ is the distance between the first two modified nodes, after which we impose the floor $\delta/20$. We then draw the conditional hypermean and site effects as above. We use this approximate fit for the diagnostic averages and estimated actions.

The stopping condition based on mass near the upper grid boundary does not bound the entire omitted tail or its moments. In six application fits, the mean $I$ under the exact grid and jitter distribution differed from quadrature by at most $2.70\times10^{-4}$. Larger differences occurred for the lower $2.5\%$ quantiles for credits: approximately $0.00108$ for $I$ and $0.000149$ for $H$. These differences exceeded the respective preset numerical tolerances of $0.001$ and $0.0001$, leading us to use quadrature for the application summaries. This comparison includes the point mass at the floor and separates deterministic approximation error from variation across finite samples of draws.

We also examined six Gaussian examples generated with fixed seeds, using $J\in\{25,100\}$, true $I\in\{.05,.75\}$, and selected profiles with $V\in\{0,2\}$. Some examples with low informativeness exceeded the numerical tolerances for diagnostic means. Keeping the generated effects and observations fixed, we compared quadrature with the limiting distribution of the grid and jitter procedure. For PM and the implemented CB estimator based on marginal posterior variances, realized RMSE changed by at most $0.244\%$ and ISEL by at most $1.185\%$ in relative terms. Their realized rank losses were unchanged. We did not repeat the full simulation, and these selected checks do not establish accuracy for all $702$ cells, nonnormal generators, GR, or every fit based on a finite number of draws. The simulation results therefore compare the procedures as implemented.

The formulas also apply to posterior draws targeting the specified model and prior. A parametric bootstrap of a frequentist fit generally targets a different distribution. Integrating over heterogeneity or using cached posterior draws does not, by itself, propagate all uncertainty in supplied first-stage inputs.

\section{Design: Proof and Implications of Proposition~4}\label{sec:E-design}

\subsection{Proof of Proposition~4}\label{subsec:E-proof}

\begin{proof}
With fixed $J$, common $\kappa>0$, and positive $n_j$,
\[
\tilde v=\frac{\kappa}{\tilde n},\qquad
\tilde n=\Big(\prod_jn_j\Big)^{1/J}\le\frac NJ.
\]
For fixed $\sigma^2>0$, $I=\sigma^2/(\sigma^2+\kappa/\tilde n)$ is strictly increasing in $\tilde n$. Equality in the arithmetic--geometric mean inequality holds exactly when all $n_j=N/J$, proving the result for continuous allocations. For positive integer allocations, transferring one unit from $n_a$ to $n_b$ whenever $n_a\ge n_b+2$ increases their product. Repeating this transfer gives site sizes that differ by at most one, provided that $N\ge J$ and the stated constraints are feasible.

Write $c=\kappa/\sigma^2>0$. Then $S(n)=n/(n+c)$ has $S''(n)=-2c/(n+c)^3<0$, so Jensen's inequality gives $\bar S\le S(N/J)$, with equality only at balance. Applying the same pairwise transfer argument to this concave function proves the integer result. Balance therefore minimizes the risk $\sigma^2(1-\bar S)$ in Proposition~2 when the inputs are known. Under the independent normal model with known hypermean, this is the Bayes risk. Under the conditional-moment assumptions alone, it is the minimum risk among affine predictors using only their own site's estimate.
\end{proof}

\subsection{Equivalent Balanced Site Size: Approximation and Population Analogue}\label{subsec:E-efficiency}

Under Proposition~4's common variance law, an unequal allocation has the same centroid as a balanced continuous allocation with the same number of sites, unchanged heterogeneity, and site size equal to the geometric mean. The ratio of geometric to arithmetic mean size gives the fraction of the average site size retained in this centroid comparison.

For a finite profile, let $\bar n=N/J$, $u_j=n_j/\bar n-1$, and $\mathrm{CV}_J^2=J^{-1}\sum_ju_j^2$. If $\max_j|u_j|\le r<1$, Taylor's theorem gives
\[
\log\frac{\tilde n}{\bar n}
=-\frac12\mathrm{CV}_J^2+R,\qquad
|R|\le\frac{J^{-1}\sum_j|u_j|^3}{3(1-r)^3}.
\]
The first-order term vanishes because $\sum_ju_j=0$. The expression $\exp(-\mathrm{CV}_J^2/2)$ therefore gives a local approximation with an explicit remainder bound. A sample CV calculated using denominator $J-1$ must first be converted to the convention used here. When deviations are not sufficiently small, profiles with the same CV can have different geometric means.

For a population analogue, suppose a positive random size satisfies $\log n\sim N(\mu_n,s_n^2)$. Then $\exp(\E\log n)=e^{\mu_n}$, $\E n=e^{\mu_n+s_n^2/2}$, and $\mathrm{CV}^2=e^{s_n^2}-1$, so
\[
\frac{\exp(\E\log n)}{\E n}
=e^{-s_n^2/2}=(1+\mathrm{CV}^2)^{-1/2}.
\]
At $\mathrm{CV}=0.75$, this population ratio is $0.80$, whereas $\exp(-0.75^2/2)\approx0.755$. The population identity neither determines the geometric mean of a realized finite profile from its CV nor makes the local approximation exact. Whether the assumed distribution of site sizes describes education trials requires empirical assessment.

\subsection{Constraints, Costs, and Alternative Targets}\label{subsec:E-variants}

Consider continuous allocations with positive feasible bounds $L_j\le n_j\le U_j$. Maximizing $\sum_j\log n_j$ subject to $\sum_jn_j=N$ gives $n_j=\min\{U_j,\max(L_j,1/\lambda)\}$, where $\lambda$ is chosen to meet the budget. With unequal positive unit costs $c_j$ and budget $\sum_jc_jn_j=B$, the unconstrained solution for the same centroid criterion is $n_j=B/(Jc_j)$. This allocation equalizes spending across sites. Both formulas optimize the stated log-product objective; other losses may favor different allocations.

With common $\kappa$, balance gives $H=V=0$ and $J_{\mathrm{eff}}=J$. If the known constants $\kappa_j$ differ but remain fixed as sizes vary, maximizing $I$ is still equivalent to maximizing the product of sizes because $\mathrm{GM}(v)=\mathrm{GM}(\kappa)/\mathrm{GM}(n)$. In this case, balance need not equalize reliabilities or maximize $\bar S$. If allocation changes the sampling-variance law itself, the objective must be recalculated.

For an independent normal population-mean problem with known $\sigma^2,v$, the GLS estimator of $\tau$ has variance
\[
\left\{\sum_j(\sigma^2+v_j)^{-1}\right\}^{-1}
=\frac{\sigma^2}{J\bar S}.
\]
This criterion also favors balance at fixed $J,N$ and common $\kappa$. Allowing the number of sites to vary, however, gives a different comparison. At balance, the variance is $\sigma^2/J+\kappa/N$, which decreases as $J$ increases, whereas $I=\sigma^2/(\sigma^2+\kappa J/N)$ decreases. Precision for the population mean, recovery of effects for a finite set of sites, and the choice of sites are therefore separate design questions \citep{BloomSpybrook2017,Miratrix2021}. Efficiency criteria in the literature on unequal cluster sizes likewise depend on the goal and model \citep{vanBreukelen2007}. The outcome ICC and population-mean formulas used there do not replace the reliability of site-effect estimates considered here.

\section{Simulation Details and Additional Results}\label{sec:F-simulation}

\subsection{Construction of Profiles with Specified \texorpdfstring{$(I, V)$}{(I, V)}}\label{subsec:F-profiles}

We construct profiles with a specified center and spread on the logit scale, then vary their shape. Informativeness and logit dispersion are fixed exactly in each stress-test cell. For specified $(I, V)$, a between-site standard deviation $\sigma = 0.20$, and $J$ sites, we set the geometric mean to $\tilde{v} = \sigma^2 (1 - I)/I$ and construct
\begin{equation}\label{eq:F-profile}
\ln v_j = \ln \tilde{v} + x_j, \qquad \frac{1}{J}\sum_j x_j = 0, \qquad \frac{1}{J}\sum_j x_j^2 = V,
\end{equation}
so the resulting $\operatorname{logit}(S_j) = \ln \sigma^2 - \ln v_j$ has mean $\operatorname{logit}(I)$ and variance $V$. We use four shapes for $\{x_j\}$:
\begin{enumerate}[label=(\alph*)]
\item \emph{Even}: $x_j$ proportional to an equally spaced grid, standardized to mean zero and mean square $V$ (sampling variances equally spaced on the log scale);
\item \emph{Bimodal}: two groups of $\lceil J/2\rceil$ and $\lfloor J/2\rfloor$ sites, obtained by centering and rescaling a two-level vector to mean zero and mean square $V$; the finite profile is asymmetric when $J$ is odd;
\item \emph{Concentrated precise}: $k = \lceil J/5 \rceil$ precise sites at $x = -\sqrt{V (J-k)/k}$ and $J - k$ diffuse sites at $x = +\sqrt{V k/(J-k)}$, giving a fifth of the sites much higher reliabilities than the remainder;
\item \emph{Concentrated noisy}: the reflection of the preceding profile, with $k$ noisy sites at $x = +\sqrt{V (J-k)/k}$ and the remainder at $x = -\sqrt{V k/(J-k)}$. We include both orientations because, at fixed $(I, V)$, the difference $\bar{S} - I$ depends jointly on the skewness of the logit profile and on $I$. This determines the direction of the dispersion effect on benchmark risk. The reflected profiles cross the $\bar{S} = I$ boundary at different informativeness levels (Appendix~\ref{sec:C-dispersion}).
\end{enumerate}
At $V = 0$ the shapes coincide, so we simulate this condition once. Spreading the $x_j$ while holding $\tilde{v}$ fixed increases total precision $\sum_j 1/v_j$ by a factor of $J^{-1}\sum_j e^{-x_j} \geq 1$. At $V = 2$ this factor is $2.35$ (even), $2.18$ (bimodal), $3.78$ (concentrated precise), and $1.63$ (concentrated noisy) for $J = 50$. The stress-test experiment therefore examines whether $I$ is sufficient to summarize performance. We compare designs with fixed resources in the fixed-enrollment experiment of Section~\ref{subsec:F-fixedN}.

We generate latent effects $\tau_j = \sigma z_j$ using the standardized generators in \textsf{multisiteDGP} (version 0.1.0): Gaussian; a two-component Gaussian mixture with separation $\delta = 5$, weight $\varepsilon = .3$, and scale ratio $\Upsilon = 2$; and asymmetric Laplace with quantile parameter $\rho = .3$. These are the three population shapes considered by \citet{LeeEtAl2025}. We then draw observed estimates as $\hat{\tau}_j \sim \mathsf{N}(\tau_j, v_j)$. Within each $(J, I, G)$ family, we draw the matrix of latent effects and the matrix of standard normal errors once and use them for all thirteen $(V, \text{shape})$ combinations. These common random numbers pair the dispersion contrasts.

\subsection{Estimation and Numerical Checks}\label{subsec:F-fitting}

For each replication, we fit the normal working model using the grid computation in Appendix~\ref{sec:D-estimation} and $M=2{,}000$ joint posterior draws. Posterior means and marginal variances are Rao--Blackwellized over the hyperparameter draws. Writing $m_j^{(r)}=\tau^{(r)}+S_j^{(r)}(\hat\tau_j-\tau^{(r)})$ and $q_j^{(r)}=\sigma^{2(r)}(1-S_j^{(r)})$, we calculate
\[
\widehat m_j=M^{-1}\sum_rm_j^{(r)},\qquad
\widehat\Sigma_{jj}=M^{-1}\sum_r\{q_j^{(r)}+(m_j^{(r)})^2\}-\widehat m_j^2.
\]
\begingroup\widowpenalty=10000
These empirical posterior moments use denominator $M$; no correction for unbiased sample variance across draws is applied. With equal $v_j$, every conditional mean has a common positive slope, and averaging over the draws preserves the ordering of the unpooled estimates. When the CB transformation below is nondegenerate, it is increasing and yields the same rank loss as PM and the unpooled estimates in a homogeneous profile. We use this algebraic equality to check the saved replications; it does not indicate superior performance.\par\endgroup

\paragraph{Constrained Bayes estimation.}
Let $m=E(\boldsymbol\tau\mid Y)$, $\Sigma=\operatorname{Cov}(\boldsymbol\tau\mid Y)$, and $P=I_J-\mathbf1\mathbf1^\top/J$. The posterior expected centered sum of squares is
\[
E(\boldsymbol\tau^\top P\boldsymbol\tau\mid Y)=m^\top Pm+\operatorname{tr}(P\Sigma).
\]
For $m^\top Pm>0$ and finite second moments, exact constrained Bayes fixes the mean at $\bar m$ and matches this centered spread, yielding
\[
d^{\mathrm{CB,exact}}=\bar m\mathbf1+
\sqrt{1+\frac{\operatorname{tr}(P\Sigma)}{m^\top Pm}}\,Pm.
\]
The posterior expected squared loss differs from $\|d-m\|^2$ by a constant. Under the fixed norm constraint, the Cauchy--Schwarz inequality gives $Pd$ proportional to $Pm$ \citep{Ghosh1992}. This solution depends on the full posterior covariance matrix.

For the simulations, we use the following CB approximation:
\[
v_m=\frac{\sum_j(\widehat m_j-\overline{\widehat m})^2}{J-1},\qquad
\overline v_{\mathrm{post}}=J^{-1}\sum_j\widehat\Sigma_{jj},
\]
\[
\widehat d_j^{\mathrm{CB}}=\overline{\widehat m}
+(\widehat m_j-\overline{\widehat m})
\sqrt{\frac{v_m+\overline v_{\mathrm{post}}}{v_m}}.
\]
Here $v_m$ uses denominator $J-1$, and we approximate $\operatorname{tr}(P\Sigma)$ by $(J-1)\overline v_{\mathrm{post}}$. The replacement is exact when $\Sigma$ is diagonal. In a hierarchical posterior, however, the shared hyperparameters generally induce off-diagonal covariances, so CB need not match the exact posterior spread of the finite ensemble. If $v_m$ is below machine epsilon, the implementation returns the common mean, which need not satisfy a positive spread constraint.

\paragraph{GR estimation, ranks, and EDF loss.}
For GR, we rank the site effects in ascending order within each joint posterior draw, assigning average ranks to ties. Averaging these ranks over the draws gives posterior expected rank scores. We order the scores, breaking exact ties by input order, and assign the $J$ quantiles of the pooled draws at $(2k-1)/(2J)$ in that order. These quantiles are obtained from the empirical inverse CDF (R quantile type1) and approximate the midpoint quantiles of
\[
\overline G(t)=E\{G_J(t)\mid Y\}=J^{-1}\sum_jP(\tau_j\le t\mid Y).
\]
For any proposed EDF $F_d$, the posterior expected ISEL is $\int(F_d-\overline G)^2$ plus a term that does not depend on the action. Among distributions assigning equal mass to $J$ points, this loss is minimized by placing the ordered coordinates at the midpoint quantiles \citep{ShenLouis1998}. Assigning site labels according to the rank scores leaves the distributional loss unchanged. Expected rank scores may take arbitrary real values and are distinct from a permutation of the sites. Both also differ from the ranks of the resulting effect estimates, with average ranks assigned to ties. The reported MSELP uses these latter ranks, divided by $J$. GR remains subject to Monte Carlo error from the posterior draws.

\paragraph{Benchmarks and loss computation.}
When $(\tau,\sigma^2)$ are known, the Gaussian linear benchmark is $\tau+S_j(\hat\tau_j-\tau)$, or $S_j\hat\tau_j$ for the zero-mean generators considered here. For non-Gaussian effect distributions, this is the affine benchmark based on the known moments and each site's own estimate; it is generally different from the Bayes rule. The semi-oracle treats $\sigma^2$ as known and estimates the hypermean by using weights $(\sigma^2+v_j)^{-1}$. We calculate losses for the realized finite collection of effects, with equal site weights. We integrate ISEL exactly between the sorted effects and estimates, where both EDFs are constant. The verification scripts check the loss calculations, rank equality for homogeneous profiles, and formal numerical examples. These checks concern numerical computation; posterior calibration across data-generating laws requires separate assessment.

\subsection{Additional Stress-Test Results}\label{subsec:F-extra}

Tables~\ref{tab:F-ratios} and \ref{tab:F-shapes} summarize paired loss ratios by exponentiating their average logarithm, with equal weight for the nine fixed $(J,G)$ families at each $I$. Table~\ref{tab:F-robust} uses the same calculation over the three $J$ families within each fixed $G$. The fixed-enrollment experiment below instead averages ratios of arithmetic cell mean losses over its design grid. These summaries estimate different quantities.

For a fixed estimator, loss, shape, and $I$, write $Z_{fr}=\log(L^{V=2}_{fr}/L^{V=0}_{fr})$, where $r=1,\ldots,R$ indexes paired replications in family $f=1,\ldots,K$. The reported geometric summary is
\[
\widehat Q=\exp(\widehat\mu),\qquad
\widehat\mu=\frac{1}{KR}\sum_{f=1}^{K}\sum_{r=1}^{R}Z_{fr},
\qquad K=9,\quad R=500.
\]
The two losses in each replication use the same latent effects and standard normal errors. Let $g_f$ denote the selected family's original identifier among the 54 families in the simulation script, as distinct from its local index $f=1,\ldots,K$. The seed for fitting is $s_0+1000g_f+61c+100000+r$, where $s_0=20260710$ and $c$ indexes the profile combination. Some combinations reuse seeds at shifted replication indices, so the fitting draws are not independent across all replications.

To estimate simulation error for the fixed grid, we join rows of paired log ratios that share a fit seed for either the numerator or the baseline. For these comparisons, each independent component contains at most three rows. We group the components by their exact multiset of family memberships, which also determines their structure under the seed patterns used here. Let $C_{tb}$ be the sum of log ratios in component $b$ of type $t$, with $B_t$ components and sample variance $s_{C,t}^2$. The estimated variance and delta-method MCSE on the ratio scale are
\[
\widehat{\operatorname{Var}}(\widehat\mu)
=\frac{\sum_t B_t s_{C,t}^2}{(KR)^2},
\qquad
\widehat{\mathrm{MCSE}}(\widehat Q)
=\widehat Q\sqrt{\widehat{\operatorname{Var}}(\widehat\mu)}.
\]
Each type contains at least 36 components. The calculation treats the data rows and streams initialized with distinct seeds as independent, as intended in the simulation design. It accounts for reuse of the same seed, although this does not constitute a formal independence result for pseudorandom bits. Variation between the fixed family means measures heterogeneity across designs and is excluded from the simulation MCSE. The component partition remains valid, though coarser than necessary, for estimators that do not use random draws in fitting.

For the $K=9$ summaries across all four shapes and the reported estimators, MCSEs on the ratio scale are at most $.00560$ for RMSE, $.00589$ for MSELP, and $.01262$ for ISEL. We do not report aggregate MCSEs for the $K=3$ summaries conditional on $G$ or the fixed-enrollment ratios. Median relative MCSEs for the arithmetic mean losses in individual cells are $.62\%$, $1.06\%$, and $2.47\%$, respectively. These errors quantify Monte Carlo precision for the specified procedures on the selected grid, without accounting for uncertainty about performance in an untested trial.

For the even profile, unpooled RMSE ratios exceed one at every reported $I$. This is consistent with the increase in integrated MSEL, $J^{-1}\sum_jv_j$, as the variance profile becomes more dispersed with its geometric mean held fixed. The PM and CB rank losses coincide because nondegenerate CB rescaling preserves PM ranks. At $I=.05$, the RMSE ratios for the even profile are $.976$ for the linear benchmark with known inputs, $.919$ for the semi-oracle with known variance, and $.863$ for fitted PM. The procedures condition on different quantities, so their differences do not provide an exact causal decomposition of the gains from the two sources of learning. For CB and GR, the corresponding ISEL ratios are approximately $.49$ and $.57$ at this $I$. These improvements refer to the chosen loss for the finite EDF and the generators used here.

The aggregate differences between shapes in Table~\ref{tab:F-shapes} do not give a universal rule for the direction of fitted-PM RMSE. Of the 648 comparisons across families, nonzero dispersion levels, and shapes, 9 give a direction different from that predicted by the $\bar S-I$ benchmark with known inputs. Comparisons across effect distributions also depend on the estimation goal (Table~\ref{tab:F-robust}). At $I=.75$, the GR ISEL ratio exceeds one for the Gaussian generator and remains below one for the mixture and asymmetric Laplace generators. These selected comparisons do not establish general robustness to distributional misspecification.

\begin{table}[!ht]
\caption{Geometric means of paired loss ratios comparing $V=2$ with $V=0$ for even profiles.}
\label{tab:F-ratios}
\centering
\begin{tablebody}
\begin{tabular}{@{}lrrrrrr@{}}
\toprule
& \multicolumn{6}{c}{Informativeness $I$} \\
\cmidrule(lr){2-7}
Estimator, loss & .05 & .15 & .30 & .45 & .60 & .75 \\
\midrule
PM, RMSE & .86 & .92 & .94 & .98 & 1.03 & 1.12 \\
\addlinespace[5pt]
CB, RMSE & .76 & .88 & .96 & 1.00 & 1.06 & 1.15 \\
\addlinespace[5pt]
GR, RMSE & .76 & .89 & .96 & 1.01 & 1.05 & 1.13 \\
\addlinespace[5pt]
ML, RMSE & 1.51 & 1.51 & 1.50 & 1.51 & 1.50 & 1.50 \\
\addlinespace[9pt]
Linear benchmark, RMSE & .98 & .95 & .96 & .98 & 1.03 & 1.11 \\
\addlinespace[5pt]
Semi-oracle, RMSE & .92 & .94 & .95 & .98 & 1.03 & 1.12 \\
\addlinespace[9pt]
PM ($=$ CB), MSELP & .93 & .92 & .94 & .99 & 1.07 & 1.20 \\
\addlinespace[5pt]
GR, MSELP & .93 & .92 & .94 & .99 & 1.06 & 1.18 \\
\addlinespace[5pt]
ML, MSELP & 1.02 & 1.05 & 1.11 & 1.19 & 1.28 & 1.40 \\
\addlinespace[9pt]
PM, ISEL & .98 & .88 & .84 & .92 & 1.05 & 1.20 \\
\addlinespace[5pt]
CB, ISEL & .49 & .76 & .83 & .89 & .96 & 1.07 \\
\addlinespace[5pt]
GR, ISEL & .57 & .77 & .76 & .82 & .88 & .98 \\
\addlinespace[5pt]
ML, ISEL & 1.07 & 1.19 & 1.34 & 1.45 & 1.52 & 1.55 \\
\bottomrule
\end{tabular}
\end{tablebody}
\par\vspace{0.4\baselineskip}
\begin{minipage}{\linewidth}\normalsize\onehalfspacing\raggedright
\textit{Note.} Entries are $\exp\{(KR)^{-1}\sum_{f,r}\log(L^{V=2}_{fr}/L^{V=0}_{fr})\}$; these differ from ratios of arithmetic mean losses. RMSE is the square root of MSEL in each replication. Families $(J,G)$ receive equal weight. The delta-method MCSE uses the components defined by shared seeds above. Its maximum across all shapes is reported in the text.
\end{minipage}
\end{table}

\begin{table}[!ht]
\caption{Geometric means of paired RMSE ratios for fitted PM by profile shape.}
\label{tab:F-shapes}
\centering
\begin{tablebody}
\begin{tabular}{@{}lrrrrrr@{}}
\toprule
& \multicolumn{6}{c}{Informativeness $I$} \\
\cmidrule(lr){2-7}
Profile shape & .05 & .15 & .30 & .45 & .60 & .75 \\
\midrule
Even & .86 & .92 & .94 & .98 & 1.03 & 1.12 \\
\addlinespace[5pt]
Bimodal & .86 & .92 & .94 & .97 & 1.04 & 1.14 \\
\addlinespace[5pt]
Concentrated precise & .84 & .94 & .99 & 1.03 & 1.08 & 1.15 \\
\addlinespace[5pt]
Concentrated noisy & .90 & .93 & .93 & .93 & .96 & 1.05 \\
\bottomrule
\end{tabular}
\end{tablebody}
\par\vspace{0.4\baselineskip}
\begin{minipage}{\linewidth}\normalsize\onehalfspacing\raggedright
\textit{Note.} Aggregation is as in Table~\ref{tab:F-ratios}. The informativeness level at which the aggregate ratio crosses one varies by shape; these crossings are not exact thresholds for fitted PM. Nine of the 648 comparisons across individual families, nonzero dispersion levels, and shapes depart from the direction predicted by the benchmark with known inputs. The maximum MCSE for the PM RMSE ratios in this table is $.00224$.
\end{minipage}
\end{table}

\begin{table}[!ht]
\caption{Geometric means of paired GR ISEL ratios by true effect distribution for even profiles.}
\label{tab:F-robust}
\centering
\begin{tablebody}
\begin{tabular}{@{}lrrrrrr@{}}
\toprule
& \multicolumn{6}{c}{Informativeness $I$} \\
\cmidrule(lr){2-7}
True $G$ & .05 & .15 & .30 & .45 & .60 & .75 \\
\midrule
Gaussian & .56 & .71 & .69 & .81 & .96 & 1.06 \\
\addlinespace[5pt]
Gaussian mixture & .64 & .79 & .78 & .79 & .82 & .89 \\
\addlinespace[5pt]
Asymmetric Laplace & .52 & .80 & .82 & .85 & .87 & .99 \\
\bottomrule
\end{tabular}
\end{tablebody}
\par\vspace{0.4\baselineskip}
\begin{minipage}{\linewidth}\normalsize\onehalfspacing\raggedright
\textit{Note.} Each entry exponentiates the average paired log ratio across replications in the three $J$ families at fixed $(I,G)$, with $R=500$ per family. This uses $K=3$ in the calculation of Table~\ref{tab:F-ratios}. The $K=9$ MCSE bounds above do not apply to these summaries conditional on $G$. At $I=.75$ the directions differ between generators, so the selected comparisons do not imply universal robustness across effect distributions.
\end{minipage}
\end{table}

\subsection{Metamodel Prediction on the Selected Grid}\label{subsec:F-fair}

Table~\ref{tab:F-fair} compares the three alternative profile summaries for all nine combinations of estimator and loss. Each model predicts the logarithm of the cell mean loss using the same factor interaction $I\times J\times G$, one summary $X\in\{H,V,\bar S\}$, and its interaction with the $I$ factor. We calculate $I$ and $X$ from the generating $\sigma^2$ and $v_j$, rather than from fitted posterior summaries. For each estimator, five-fold cross-validation randomly partitions the rows of cell means and predicts the held-out rows of the selected 702-cell grid. Other profiles from the same families, sharing common random numbers, remain in the training data. We do not assess prediction for held-out effect-distribution generators or external trials. The $R^2$ values therefore describe prediction within this grid, and the residual variation cannot be attributed entirely to Monte Carlo error.

\begin{table}[!ht]
\caption{Five-fold cross-validation $R^2$ across cell means for all combinations of estimator and loss.}
\label{tab:F-fair}
\centering
\begin{tablebody}
\begin{tabular}{@{}llrrr@{}}
\toprule
Loss & Estimator & $H$ & $V$ & $\bar S$ \\
\midrule
RMSE & PM & .98964 & .99155 & .99843 \\
\addlinespace[5pt]
RMSE & CB & .99099 & .99336 & .99816 \\
\addlinespace[5pt]
RMSE & GR & .99076 & .99308 & .99784 \\
\addlinespace[9pt]
MSELP & PM & .99269 & .99452 & .99923 \\
\addlinespace[5pt]
MSELP & CB & .99269 & .99452 & .99923 \\
\addlinespace[5pt]
MSELP & GR & .99298 & .99470 & .99919 \\
\addlinespace[9pt]
ISEL & PM & .99598 & .99663 & .99757 \\
\addlinespace[5pt]
ISEL & CB & .99389 & .99575 & .99693 \\
\addlinespace[5pt]
ISEL & GR & .99586 & .99572 & .99518 \\
\bottomrule
\end{tabular}
\end{tablebody}
\par\vspace{0.4\baselineskip}
\begin{minipage}{\linewidth}\normalsize\onehalfspacing\raggedright
\textit{Note.} $I$, $H$, $V$, and $\bar S$ are calculated from the true generating profiles. All models include the same interaction among factors, together with an interaction between the profile summary and $I$. Cross-validation randomly partitions rows of cell means within the selected simulation grid, so the results do not validate prediction for new families or trials. Values are $R^2$ for predicting the logarithm of mean loss in held-out rows. These prediction comparisons do not quantify confidence or calibration for site effects, ranks, or EDFs.
\end{minipage}
\end{table}

\subsection{Fixed-Enrollment Experiment}\label{subsec:F-fixedN}

The fixed-enrollment experiment holds $N=J\bar n$ constant and uses $v_j=4/n_j$. We compare ML, PM, CB, GR, and the linear benchmark with known $(\tau,\sigma^2)$. For each $(J,\bar n,G)$ family, we draw one Gamma vector with parent mean $\bar n$ at each positive value of the parent distribution's coefficient of variation $\mathrm{CV}_{\Gamma}$. Values below 5 are set to 5, and the vector is rescaled to sum to $N$; both operations are then repeated. The resulting sizes are continuous design quantities. They are neither integer enrollments nor draws from a Gamma distribution truncated below 5. Some entries fall slightly below 5 after the final rescaling: the reconstructed minimum is $4.9908$, and 14 of the 108 profiles contain at least one size below 5. The parent $\mathrm{CV}_{\Gamma}$ controls the construction but differs from the realized coefficient of variation of the transformed profile. These illustrative levels have not been calibrated to empirical educational trials.

Pooled loss ratios increase across the CV levels of the parent distribution (Table~\ref{tab:F-fixedN}), although the pattern does not hold for every sequence in the grid. Of the 405 sequences across 27 families, five procedures, and three losses, 41 are nonmonotonic, and 14 contain at least one dispersed design with mean loss below its balanced counterpart. Proposition~4 assumes known inputs and a common variance law inversely proportional to site size. Its allocation result for the specified criterion does not imply monotonicity for every fitted estimator and loss considered here.

\begin{table}[!ht]
\caption{Fixed-enrollment experiment: arithmetic means of ratios of cell mean losses to their balanced counterparts.}
\label{tab:F-fixedN}
\centering
\begin{tablebody}
\begin{tabular}{@{}lrrrrrrrr@{}}
\toprule
& \multicolumn{3}{c}{PM} & \multicolumn{3}{c}{GR} & \multicolumn{2}{c}{Mean diagnostics} \\
\cmidrule(lr){2-4}\cmidrule(lr){5-7}\cmidrule(lr){8-9}
$\mathrm{CV}_{\Gamma}$ & RMSE & MSELP & ISEL & RMSE & MSELP & ISEL & $I$ & $V$ \\
\midrule
$.25$ & 1.00 & 1.01 & 1.01 & 1.01 & 1.01 & 1.01 & .327 & .06 \\
\addlinespace[5pt]
$.50$ & 1.02 & 1.04 & 1.05 & 1.03 & 1.04 & 1.03 & .308 & .27 \\
\addlinespace[5pt]
$.75$ & 1.03 & 1.08 & 1.11 & 1.05 & 1.08 & 1.05 & .279 & .61 \\
\bottomrule
\end{tabular}
\end{tablebody}
\par\vspace{0.4\baselineskip}
\begin{minipage}{\linewidth}\normalsize\onehalfspacing\raggedright
\textit{Note.} The arithmetic mean loss in each cell is divided by its balanced counterpart, and the resulting ratios are averaged with equal weight over the 27 $(J,\bar n,G)$ families: $J\in\{25,50,100\}$, $\bar n\in\{20,50,100\}$, and three generators. CV levels within a family use common data draws. Diagnostics are also arithmetic averages over the grid. At balance, their means are $I=.333$, $V=0$, whereas individual balanced $I$ values depend on $\bar n$. The parent Gamma CV differs from the realized CV of the finite profile.
\end{minipage}
\end{table}

\subsection{Computation and Numerical Accuracy}\label{subsec:F-repro}

The simulation results use archived losses for each replication. We reproduced the reported stress-test and fixed-enrollment results by recomputing arithmetic cell means, then recomputed the metamodels and five-fold splits from those means. We retained the paired loss-ratio summaries, recalculated their simulation errors using the shared-seed components above, and reran the numerical verification suite with its original assertions. We did not rerun the full historical simulation, whose runtime remains incompletely documented. The unused Jensen-gap illustration and superseded robustness summary have unresolved provenance.

All fitted procedures in the archived simulations use the stated 250-node grid, jitter, and $M=2{,}000$ draws. For selected cases, we compared the posterior sampler's distribution with adaptive quadrature under the same likelihood and reference prior flat in the mean and SD. In six additional Gaussian examples near the boundary, several cases with low informativeness gave diagnostic discrepancies exceeding the strict reporting tolerances. Comparisons of the PM and implemented CB estimates and losses quantify these selected differences, without validating all 702 stress-test cells or GR. The simulation results therefore describe the performance of the specified approximate procedures; exact Bayes performance has not been established over the full grid. The application summaries use the separate numerical assessment described in Appendix~\ref{sec:D-estimation}.

\section{Details of the Online Credit Recovery Study Analysis}\label{sec:G-ocrs}

\subsection{Data and Site Definition}\label{subsec:G-data}

The public-use OCRS impact-analysis file contains $1{,}737$ records for student--subject combinations in the 2018 and 2019 summer terms \citep{AIR2024, Rickles2024}. These represent $1{,}683$ students, including $54$ who appear in both subjects, with no duplicate student--subject records. We define sites as school--subject cells, combining the two English 9 semester courses within each school. The $24$ schools give $32$ sites ($13$ Algebra, $19$ English), with eight schools contributing both subjects. Site sizes range from $24$ to $104$ records (median $49$), and arm sizes range from $8$ to $53$. The $144$ randomization blocks within sites are defined by school, subject, cohort, and, in some schools, the semester of course failure. Twenty-eight sites contain multiple blocks (up to $10$), and six include both summer terms.

\subsection{Outcomes}\label{subsec:G-outcomes}

Course passing (\texttt{scm\_passcrs}) and subject-matched year-four credits (\texttt{cby\_y4\_math} for Algebra, \texttt{cby\_y4\_eng} for English) are observed for all analysis records. We use the released coding of course passing, under which grades of D or better count as passing. Subject-matched credits range from $0$ to $90$ in Algebra sites and from $0$ to $85$ in English sites, in district credit units. These outcomes avoid the nonresponse problem of the end-of-course test, which is missing for $38.8\%$ of records. Their completeness does not, however, eliminate uncertainty in the estimated sampling variances.

\subsection{Block-Adjusted Site-Level Estimation}\label{subsec:G-estimation}

Within each site, we obtain $\hat\tau_j$ as the treatment coefficient from an ordinary least squares regression on assignment and block indicators. Suppressing the site subscript, let $n_k$ denote block size, $p_k=n_{1k}/n_k$ the treated proportion, and $\mathcal K_{01}=\{k:0<p_k<1\}$ the set of informative blocks. Provided that the total residualized treatment variation is positive, removing block means gives
\begin{equation}\label{eq:G-block-weights}
 \hat\tau_{\mathrm{FE}}
 =\sum_{k\in\mathcal K_{01}}\omega_k(\bar Y_{1k}-\bar Y_{0k}),
 \qquad
 \omega_k=\frac{n_kp_k(1-p_k)}
 {\sum_{\ell\in\mathcal K_{01}}n_\ell p_\ell(1-p_\ell)}.
\end{equation}
Under complete randomization with fixed arm counts within informative blocks, this coefficient estimates the average of block effects with these weights. The average effect over all students uses weights $n_k/\sum_\ell n_\ell$ \citep{PashleyMiratrix2021}. For arbitrary heterogeneous block effects, the two weighting definitions agree only when every block is informative and $p_k(1-p_k)$ is common to all blocks. When block effects are equal, the two averages can coincide despite different weights.

In the analysis file, $122$ blocks are informative and $22$ contain only one arm. The latter account for $32$ records in $17$ sites. A further $27$ informative blocks have only one record in an arm. The regression weights differ from weights for the average over all students in $28$ of the $32$ sites; the within-site sum of absolute weight differences has median $.0739$ and maximum $.3383$. We retain the regression contrast as the first-stage target. Blocks containing only one arm supply no treatment contrast of their own, and the small cells prevent uniform application of a variance formula based on arm sample variances. The observed regression calculation remains valid, although these features limit its interpretation as an average effect over all students.

For a retained design basis $X$ of full rank, residuals $e_i$, and leverages $h_i$, we calculate the HC2 covariance \citep{MacKinnonWhite1985}, imposing a lower bound on the denominator:
\begin{equation}\label{eq:G-hc2}
 \widehat{\operatorname{Var}}(\hat\beta)
 =(X^\top X)^{-1}X^\top
 \operatorname{diag}\!\left\{\frac{e_i^2}{\max(1-h_i,10^{-8})}\right\}
 X(X^\top X)^{-1}.
\end{equation}
The diagonal element for the treatment coefficient gives $\hat v_j$. The lower bound on the denominator is active in $15$ rows of each outcome fit with block indicators only and $17$ rows of the covariate-adjusted fit. Leverage is numerically one in these rows. Their treatment-score coefficients, from the treatment row of $(X^\top X)^{-1}X^\top$, have absolute values below $6\times10^{-16}$. Their contribution to the estimated treatment variance is below $3\times10^{-49}$ of the total in the original and bootstrap fits. The lower bound thus handles numerically degenerate rows with essentially zero treatment scores in this application; it provides no general correction to coverage. Each bootstrap first stage has $15$ safeguarded rows because cell counts and the treatment design are held fixed.

The design matrices with block indicators only are of full rank. Covariate adjustment adds the released imputed subject-specific grade-8 test score and ninth-grade GPA, together with missingness indicators. In $30$ sites, the GPA missingness indicator is redundant and is dropped by the QR decomposition; the treatment coefficient remains identified. The weights in \eqref{eq:G-block-weights} do not describe the regression with these additional covariates. Separate calculations using residualized treatment reproduce the reported coefficients and variances, and all supplied variances are positive. Differences in treated proportions between terms reach $.3125$ (median $.0810$ among the six sites spanning two terms). Block adjustment changes course passing estimates by at most $.243$ (median $.029$).

HC2 adjusts for leverage, but does not remove estimation uncertainty in $\hat v_j$ or guarantee exact finite-sample inference under heteroskedasticity or with small binary cells. In the second-stage working likelihood, we set $v_j=\hat v_j$ and interpret the posterior conditionally on these inputs.

\subsection{Model Fitting and Diagnostics}\label{subsec:G-fitting}

We fit the independent normal--normal working model with reference measure $d\tau\,d\sigma$, which is flat in the mean and SD. We compute posterior means and intervals for the functionals by adaptive integration over the untruncated marginal distribution of $\sigma$, using the tail and numerical checks in Appendix~\ref{subsec:D-algorithm}. A flat reference for the SD differs from a flat prior on the variance. Under the independent working likelihood with fixed positive inputs, the site counts satisfy the posterior propriety and moment conditions in Appendix~\ref{subsec:D-prior}. These conditions establish that the posterior quantities exist, but not that the model is adequate.

Table~\ref{tab:ocrs} gives the profile summaries. Table~\ref{tab:G-ocrs-ancillary} adds the fitted population mean, heterogeneity SD, plug-in comparison, posterior probability that informativeness exceeds one half, and $V$. Informativeness exceeds one half exactly when the signal variance exceeds the geometric-mean sampling variance. The plug-in statistic is $I\{\E(\sigma^2\mid\text{data})\}$, whereas we report $\E\{I(\sigma^2)\mid\text{data}\}$ for informativeness. As Proposition~3 implies, the plug-in value is larger in every fit. Conditional on the supplied variances, $V=\operatorname{Var}_j(\log v_j)$ is constant across posterior draws, although it can change when the first stage is resampled. The ratios of largest to smallest variance are $7.9$ for pooled course passing, $8.9$ for adjusted course passing, and $10.4$ for credits.

\begin{table}[!t]
\caption{Additional OCRS posterior summaries.}
\label{tab:G-ocrs-ancillary}
\centering
\begin{tablebody}
\begin{tabular}{@{}lccccc@{}}
\toprule
Analysis & $\tau$ & $\sigma$ [95\% CrI] & $I_{\mathrm{plug}}$ & $\Pr(I>\tfrac12)$ & $V$ \\
\midrule
Course passing & $-.130$ & $.188\ [.127,.266]$ & $.691$ & $.973$ & $.237$ \\
\addlinespace[5pt]
Algebra 1 & $-.081$ & $.189\ [.086,.339]$ & $.691$ & $.818$ & $.158$ \\
\addlinespace[5pt]
English 9 & $-.163$ & $.203\ [.122,.318]$ & $.740$ & $.972$ & $.282$ \\
\addlinespace[5pt]
Adjusted & $-.117$ & $.184\ [.124,.261]$ & $.673$ & $.955$ & $.260$ \\
\addlinespace[9pt]
Credits & $-.124$ & $2.807\ [.610,4.949]$ & $.353$ & $.114$ & $.313$ \\
\bottomrule
\end{tabular}
\end{tablebody}
\par\vspace{0.4\baselineskip}
\begin{minipage}{\linewidth}\normalsize\onehalfspacing\raggedright
\textit{Note.} The first four rows concern course passing; credits are matched to subject. Entries for $\tau$ and $\sigma$ are posterior means for the population of site effects under the working model, with an interval for $\sigma$. Here $I_{\mathrm{plug}}=I\{\E(\sigma^2\mid\text{data})\}$. The probability column also conditions on the data and supplied variances, under which $V$ is fixed. Both $\tau$ and $\sigma$ are expressed in outcome units, so their magnitudes are not directly comparable between course passing and credits.
\end{minipage}
\end{table}

The subject-specific fits describe different finite site collections, and course passing and credits differ in outcome and units. The profile diagnostics are invariant to a common linear change of units, but this does not account for differences in scientific target. For binary risk differences, the normal first-stage likelihood remains an approximation. Neither the minimum site sizes nor similarity across adjusted fits establishes its coverage.

\subsection{Sensitivity Analyses}\label{subsec:G-sensitivity}

\emph{One site per school.} We refit all $2^8=256$ selections retaining one subject per paired school, each with $J=24$. For course passing, the posterior mean of $I$ obtained by integration ranges from $.612$ to $.748$ (median $.690$); for credits, the posterior mean of $I$ ranges from $.116$ to $.421$ (median $.266$). These means remain above one half for course passing and below it for credits in every collection. Each selection changes the set of sites, subject composition, and equally weighted finite profile. The ranges therefore describe sensitivity to the site collection; they do not describe joint covariance or interval coverage for the original $32$ sites. All $54$ students observed twice attend the same school in both subjects. Their assignment arms agree in $37$ cases and differ in $17$. Neither site selection nor the independent likelihood reconstructs the full sampling covariance due to this overlap.

\emph{Effect scale and specification.} The unadjusted risk difference is the difference between course passing proportions in the two arms of a site. Its variance is the sum of the two arm sample variances divided by their respective arm sizes. We calculate the log odds ratio from the same $2\times2$ table, with the Woolf reciprocal-count variance. The prescribed correction adds $0.5$ to all four cells when any cell is zero, but is inactive in every site. Table~\ref{tab:G-ocrs-scale} shows separately the effect of block adjustment and the difference between the two unadjusted effect measures. Both comparisons change informativeness. Even under matching specifications, risk differences and log odds are not related by a common linear rescaling, because their relation depends on baseline risks. Both between-site signal and sampling noise change, so a reduction in effect spread alone cannot account for the differences. The primary illustration uses block-adjusted risk differences.

\begin{table}[!t]
\caption{Course passing diagnostics by first-stage specification and effect scale.}
\label{tab:G-ocrs-scale}
\centering
\begin{tablebody}
\begin{tabular}{@{}lcccc@{}}
\toprule
Specification & $I$ [95\% CrI] & $\Pr(I>\tfrac12)$ & $H$ & $V$ \\
\midrule
Block-adjusted RD & $.671\ [.497,.812]$ & $.973$ & $.025$ & $.237$ \\
\addlinespace[5pt]
Unadjusted RD & $.632\ [.439,.790]$ & $.920$ & $.021$ & $.188$ \\
\addlinespace[5pt]
Unadjusted log OR & $.530\ [.253,.743]$ & $.627$ & $.035$ & $.306$ \\
\bottomrule
\end{tabular}
\end{tablebody}
\par\vspace{0.4\baselineskip}
\begin{minipage}{\linewidth}\normalsize\onehalfspacing\raggedright
\textit{Note.} All fits use $J=32$ sites and the normal working model with a reference measure flat in the mean and SD. Entries for $I$ and $H$ are posterior means obtained by integration. RD denotes risk difference and OR odds ratio. The unadjusted rows use the same counts by site and arm; the block-adjusted row uses the regression contrast in \eqref{eq:G-block-weights}. Credible intervals condition on the variances supplied for each specification.
\end{minipage}
\end{table}

\emph{First-stage resampling and variance floors.} We hold the original sites and the counts in each combination of site, arm, and block fixed, and resample student records independently with replacement within each cell. We recompute the block-adjusted first stage and hierarchical fit for $B=500$ samples per outcome. For each refit, we compute the posterior means $I_b$ and $H_b$ by integration; $V_b$ is the centered spread of the supplied log variances. Table~\ref{tab:G-ocrs-bootstrap} gives the empirical .025 and .975 quantiles of these $500$ summaries. The quantiles are calculated across refitted summaries, without pooling posterior draws, and are distinct from credible intervals conditional on the original data.

Some binary resamples yield HC2 variances close to zero. We therefore calculate a fixed floor before resampling from the original arm counts $n_{ja}$:
\begin{equation}\label{eq:G-variance-floor}
 p_{\min,ja}=\frac{0.5}{n_{ja}+1},\qquad
 v_{\min,j}=\sum_{a=0}^1\frac{p_{\min,ja}(1-p_{\min,ja})}{n_{ja}}.
\end{equation}
This floor uses the extreme-count continuity correction for zero successes, without substituting observed or resampled success counts. For credits, the floor is $.05$ times the original site's HC2 variance. Each refit uses $v_{bj}=\max(\hat v_{bj},v_{\min,j})$. The course passing floor is active in $110$ of $16{,}000$ site refits ($.6875\%$), with one site affected in each of $110$ bootstrap samples. It is never active for credits.

\begin{table}[!t]
\caption{Central 95\% spreads of refitted summaries for alternative variance floors.}
\label{tab:G-ocrs-bootstrap}
\centering
\begin{tablebody}
\begin{tabular}{@{}lccc@{}}
\toprule
Outcome (floor factor) & $I_b$ & $H_b$ & $V_b$ \\
\midrule
Course passing ($0.5\times$) & $[.665,.828]$ & $[.016,.032]$ & $[.191,.353]$ \\
\addlinespace[5pt]
Course passing ($1\times$) & $[.665,.828]$ & $[.015,.031]$ & $[.191,.324]$ \\
\addlinespace[5pt]
Course passing ($2\times$) & $[.662,.825]$ & $[.015,.030]$ & $[.188,.318]$ \\
\addlinespace[9pt]
Credits ($0.5$--$2\times$) & $[.377,.726]$ & $[.025,.052]$ & $[.234,.468]$ \\
\bottomrule
\end{tabular}
\end{tablebody}
\par\vspace{0.4\baselineskip}
\begin{minipage}{\linewidth}\normalsize\onehalfspacing\raggedright
\textit{Note.} Each row uses the same 500 bootstrap samples per outcome, obtained by resampling within cells, and reports empirical .025/.975 quantiles. Here $I_b$ and $H_b$ are posterior means of the functionals obtained by integration for refit $b$, and $V_b$ is its log-variance spread. Factor $1$ denotes the original floor; credits is unchanged for factors $0.5$, $1$, and $2$. These spreads across refits are distinct from posterior credible intervals conditional on the original data.
\end{minipage}
\end{table}

Using the same resamples, we also halve and double the floors. The mean of course passing $I_b$ changes by $+.000391$ and $-.001041$, respectively, and the largest absolute change in a single refit is $.00518$ or $.00843$. The upper percentile of $V_b$ is more sensitive: it increases from $.324$ to $.353$ when the floor is halved. Credits remains unchanged even when its floor is doubled. The sensitivity analysis is confined to this range of floors and does not cover all possible regularization rules.

The mean of $I_b$ across refits is $.762$ for course passing and $.569$ for credits, compared with original posterior means of $.671$ and $.318$. Resampling and refitting therefore change the location as well as the spread of the summaries, especially for credits. The floor cannot explain the shift for credits, where it is never active. The shift shows sensitivity to this resampling and refitting scheme; its source remains unresolved. The scheme does not jointly resample records of students who appear in both subjects or resample schools as clusters, leaving dependence due to shared units unresolved. The refit spreads do not validate the original conditional posterior intervals or establish repeated-sampling coverage for the latent reliability profile.

\subsection{Software and Replication}\label{subsec:G-replication}

The R package \textsf{poolinfo} (version~0.1.0) implements the reliability-profile diagnostics developed here. The function \texttt{reliability\_\allowbreak profile()} calculates diagnostics from standard errors and a specified between-site standard deviation. Given site-level estimates and standard errors, \texttt{reliability\_\allowbreak posterior()} integrates posterior functionals under the normal working model. The default analysis assumes a common mean and a reference measure flat in the mean and SD, conditional on the supplied standard errors. 

\vspace{0.5\baselineskip}
\noindent\begin{minipage}{\linewidth}
The following example uses the aggregated OCRS course passing estimates bundled with the package:
\begin{lstlisting}[language=R]
library(poolinfo)
fit <- reliability_posterior(
  estimate, se = se, site = site, data = ocrs,
  scale_label = "risk difference")
profile_summary(fit)
site_table(fit)
plot(fit)
\end{lstlisting}
\end{minipage}\par

The vignettes describe profile interpretation, posterior summaries, design calculations, and the OCRS application. The implementation vignette gives details of the integration methods and numerical checks. Appendix~\ref{subsec:D-algorithm} describes the reference computation used in this manuscript. The package can be applied to new site summaries. The study-specific replication scripts contain the code for the simulations, first-stage estimation, and sensitivity analyses reported here. The data and software statement lists links to the package source, documentation, and study replication repository.

The public data release is available from openICPSR project 202181 \citep{AIR2024}. The analysis script \texttt{05\_ocrs\_audit.R} reconstructs the site inputs and design checks, retains the subset selections and aggregate bootstrap inputs, and produces the scale and floor comparisons. The shared script \texttt{05\_posterior\_stability.R} provides the integration routine and numerical checks. We retain random seeds and refit summaries for reproducibility. The tables use these outputs, and the figure code uses the corresponding corrected profile summaries and posterior representation. Reproducing the numerical results does not itself validate the causal target, working likelihood, or repeated-sampling coverage.

\end{document}